%% file: main.tex
\documentclass[sigconf,prologue,table,xcdraw]{acmart}

\copyrightyear{2023} 
\acmYear{2023} 
\setcopyright{rightsretained} 
\acmConference[SIGSPATIAL '23]{The 31st ACM International Conference on Advances in Geographic Information Systems}{November 13--16, 2023}{Hamburg, Germany}
\acmBooktitle{The 31st ACM International Conference on Advances in Geographic Information Systems (SIGSPATIAL '23), November 13--16, 2023, Hamburg, Germany}\acmDOI{10.1145/3589132.3625661}
\acmISBN{979-8-4007-0168-9/23/11}

\usepackage{subcaption}
\usepackage{tikz}
\usepackage{pgfplots}

\begin{document}


\title[Unveiling limitations of synthetic mobility data]{Reconsidering utility: unveiling the limitations of synthetic mobility data generation algorithms in real-life scenarios}

\author{Alexandra Kapp}
\email{alexandra.kapp@htw-berlin.de}
\orcid{0000-0002-8348-8958}
\affiliation{%
  \institution{Hochschule für Technik und Wirtschaft Berlin, University of Applied Sciences}
  \city{Berlin}
  \postcode{12459}
  \country{Germany}
}

\author{Helena Mihaljević}
\email{helena.mihaljevic@htw-berlin.de}
\orcid{0000-0003-0782-5382}
\affiliation{%
  \institution{Hochschule für Technik und Wirtschaft Berlin, University of Applied Sciences}
  \streetaddress{Wilhelminenhofstraße 75A}
  \city{Berlin}
    \postcode{12459}
  \country{Germany}}


\begin{abstract}

In recent years, there has been a surge in the development of models for the generation of synthetic mobility data. These models aim to facilitate the sharing of data while safeguarding privacy, all while ensuring high utility and flexibility regarding potential applications. However, current utility evaluation methods fail to fully account for real-life requirements.

We evaluate the utility of five state-of-the-art synthesis approaches, each with and without the incorporation of differential privacy (DP) guarantees, in terms of real-world applicability. Specifically, we focus on so-called trip data that encode fine granular urban movements such as GPS-tracked taxi rides. 
Such data prove particularly valuable for downstream tasks at the road network level. Thus, our initial step involves appropriately map matching the synthetic data and subsequently comparing the resulting trips with those generated by the routing algorithm implemented in OpenStreetMap, which serves as an efficient and privacy-friendly baseline.

 Out of the five evaluated models, one fails to produce data within reasonable computation time and another generates too many jumps to meet the requirements for map matching.
 The remaining three models succeed to a certain degree in maintaining spatial distribution, one even with DP guarantees.
 However, all models struggle to produce  meaningful sequences of geo-locations with reasonable trip lengths and to model traffic flow at intersections accurately.
 It is important to note that trip data encompasses various relevant characteristics beyond spatial distribution, such as temporal information, all of which are discarded by these models. Consequently, our results imply that current synthesis models fall short in their promise of high utility and flexibility. 
\end{abstract}


\received{20 February 2007}
\received[revised]{12 March 2009}
\received[accepted]{5 June 2009}

\maketitle

\section{Introduction}
The field of synthetic mobility data generation has experienced rapid growth, primarily fueled by privacy concerns that hinder the release of sensitive personal movement data. In essence, respective algorithms learn statistical distributions from raw data and employ them to generate synthetic data that emulate a similar distribution, while (supposedly) safeguarding  sensitive personal information. Within the domain of mobility data, a variety of algorithmic approaches exist that can be categorized in terms of the specific type of mobility they aim to model. 
A frequently considered type of mobility in respective literature refers to `trips'  which consist of fine granular routes connecting an origin and a destination, such as taxi rides or GPS-tracked bicycle tours. This paper is focused on such trips, in contrast to mobility data that entails per person a sequence of stay locations over the course of a longer period of time, such as check-ins at restaurants or other points of interest. 

One primary objective of trip data generation algorithms is to produce `plausible' trips. 
Respective evaluations typically involve comparing aggregate statistics between raw and synthetic data, such as spatial distributions where space is commonly discretized based on a grid of a selected resolution. 
This operationalization often falls short in meeting real-life requirements: depending on the underlying (coarse) grid, the resulting synthetic trips usually do not follow the road network and instead jump over buildings and rivers in an unrealistic manner \cite{sun_synthesizing_2023}.

However, the benefit of trip data, unlike other types of mobility data, is its fine granularity such that it can be mapped to a road network. This especially enables analyses such as calculating average speeds or traffic volumes per road segment, which are valuable for applications like routing and urban planning. Furthermore, accurate representation of traffic flow at intersections facilitates optimized traffic light management \cite{wei_survey_2020}. Bicycle trips, collected by smartphone apps like SimRa~\cite{karakaya_simra_2020-1}, enable the identification of road traffic hazard zones.  

Therefore, we evaluate state-of-the-art synthesis algorithms by initially matching their generated trips to the road network. Subsequently, we compare the matched trips with those generated by the routing engine in OpenStreetMap which is a simple, efficient, and privacy-sensitive way to produce road-level movement data based on origin-destination (OD) pairs. We argue that synthetic data generation algorithms need to provide a higher utility than common routing engines to justifiably provide added value.

In summary, we address the following research questions: 

\begin{itemize}
    \item (RQ1) What constitutes high utility of trip data and how can it be measured?
    \item (RQ2) What level of utility do state-of-the-art synthesis models reach on these measures in comparison to a routing baseline?
    \item (RQ3) Can a satisfactory utility level still be provided given (differential) privacy guarantees?
\end{itemize}

The remainder of this paper is structured as follows: We commence by presenting an overview of the five evaluated state-of-the-art synthesis algorithms. In Section \ref{sec:def-utility}, 
we address RQ1 and propose a formulation of utility for trip data and suitable approaches for its measurement. Section \ref{sec:exSetup} focuses on the experimental setup, while the results are presented in Section \ref{sec:results}. Finally, we conclude by discussing our findings and possible future research.

\section{Synthesis Algorithms}
\label{sec:algorithms}

In recent years, a series of models for generating synthetic mobility data have emerged, aiming to facilitate the sharing of fine-granular datasets in a privacy-preserving manner. These  algorithms  learn  statistical distributions from a raw dataset and generate a synthetic counterpart based on the learned distributions. 
Nonetheless, without additional privacy measures, there is no guarantee that the synthesis algorithm does not inadvertently reproduce real trips or reveal sensitive information through the preserved statistical distributions \cite{luca_survey_2021}. To address this concern, many models incorporate additional privacy mechanisms, often in the form of \textit{Differential Privacy (DP)} guarantees. 

DP provides mathematical guarantees for preserving individual privacy \cite{dwork_differential_2006}. The core idea behind  DP is that the output of an algorithm remains largely unaffected if the records of a single individual are either removed or added. This mechanism effectively limits the influence of a single individual on the overall analysis outcome, thus preventing the reconstruction of their specific data. Generally, a typical approach to achieving DP in numeric functions consists of adding  calibrated noise drawn from a Laplace distribution to the function’s output. In the context of synthesis algorithms, noise is commonly added to the underlying distributions such as that of origins or OD pairs before synthetic samples are generated.

Amongst the published synthesis algorithms, we have carefully chosen five models for our evaluation to ensure a diverse representation of modeling techniques. Our selection process took into consideration various factors, especially the clarity of reasoning behind their approaches, promising results demonstrated in respective evaluations, and the availability of source code, either as open source or obtainable upon request. The selection process resulted in the following models:  \textit{AdaTrace}\cite{gursoy_utility-aware_2018}, \textit{PrivTrace}\cite{wang2023privtrace}, \textit{BiLSTM}\cite{blanco-justicia_generation_2022}, \textit{DP-Loc}\cite{lestyan_search_2022}, and \textit{TrajGAIL}\cite{choi_trajgail_2021}. 

\textbf{AdaTrace}
\cite{gursoy_utility-aware_2018} is a frequently cited and benchmarked differentially private model. 
Coordinates are discretized using an $NxN$ grid of uniform cells. Synthetic trips are generated in three steps: First, OD pairs are sampled from a differentially private OD distribution. Second, the number of points per sequence is sampled from a differentially private distribution of the respective OD pair. 
Third, a sequence between a start and an endpoint is constructed by repeatedly sampling the next location until the sequence length is reached. The next-location sampling is based on a Markov model which holds DP transition probabilities for each location, i.e., grid cell, to each other location, given the previous location and the destination. The exact coordinates of the synthetic locations are sampled uniformly at random from within the grid cell.

To ensure a more accurate spatial distribution while maintaining a high level of privacy, the authors make use of a `density-aware grid'. When a grid cell contains a small number of records, it is retained at a coarse top-level resolution. On the other hand, when the number of records in a grid cell is large, it is split into a finer-grained resolution. Note that the Markov model relies on the coarse top-level grid, with the density-aware grid coming into play only during the final step of sampling coordinates. This comes with some implications for the resulting synthetic data: Despite using a coarse top-level grid, the aggregate spatial distribution is well maintained due to the preservation of hotspots through the density-aware grid (see Appendix B of \cite{gursoy_utility-aware_2018}). However, it is worth noting that the distance between two consecutive points in the generated data is determined by the resolution of the coarse top-level grid. 
The compatibility with DP is an essential aspect of the model's architecture, and therefore, the implementation does not foresee the possibility of training the model without DP.


The recently proposed model \textbf{PrivTrace} \cite{wang2023privtrace} aims to improve upon AdaTrace, which supposedly lacks sufficient transition information due to its first-order Markov chain model. Additionally, it addresses the limitations of DPT \cite{he_dpt_2015}, where the  high-order Markov chain model introduces excessive noise due to the incorporation of DP.
PrivTrace follows a similar approach as AdaTrace but with some notable distinctions. In PrivTrace, synthetic sequences are generated using a first-order and a second-order Markov chain. Sequence sampling is stopped based on `virtual ends' in the Markov model instead of a sampled sequence length and a drawn destination. 
Additionally, PrivTrace incorporates transition probabilities of more fine-granular grid cells into the Markov Model, claiming to thereby provide better transition information. Thus, unlike AdaTrace, the distance between consecutive points is \textit{not} determined by the top-level grid resolution. Again, this model always provides DP guarantees.

Lesty{\'a}n et al. \cite{lestyan_search_2022} propose \textbf{DP-Loc}, a differentially private model that initially performs a dimensionality reduction by only considering cells that have been visited more often than a certain threshold. Then, similar to AdaTrace, DP-Loc first generates OD pairs and then constructs a trip based on transition probabilities.  For the OD generation, a variational autoencoder is used, while transition probabilities are captured using a feedforward neural network (FNN).
The authors argue that the basic FNN is adequate, as the information of the current location alone is sufficient to predict the next location, unlike, for example, models that specifically aim to model sequences.
To guarantee DP, Gaussian noise is added to the distribution of visit counts per cell \cite{8187424}, and both neural networks are trained using Differentially Private Stochastic Gradient
Descent (DP-SGD) \cite{abadi_deep_2016}, which adds noise to clipped gradients during training.
Notably, unlike the two other models, DP-Loc includes the generation of start times of trips.

Blanco-Justicia et al. \cite{blanco-justicia_generation_2022} propose the utilization of a \textbf{BiLSTM}, a bidirectional long short-term memory network, a recurrent architecture that has proven superior performance in modeling sequences such as time series or natural language. Considering a visited location as a word and a trip as a sentence, the principles of autoregressive text generation using RNNs can be adopted to generate trips.   For the task of trip generation, a start location is selected at random. Using the BiLSTM, the location is extended in an autoregressive manner to generate subsequent locations until the sequence length sampled from the respective distribution is reached.
The authors propose a  privacy mechanism that samples one of the top three next-location predictions uniformly at random from the probability distribution of the next locations.
This approach comes with a few downsides. Firstly, privacy is not formally guaranteed.
Second, utility is not well maintained, as the  results in the original paper indicate major jumps between consecutive points.
We thus follow the authors' advice for future work and apply a DP mechanism to the model using DP-SGD, more specifically, the Tensorflow implementation of a DP Adam optimizer.

\textbf{TrajGAIL} \cite{choi_trajgail_2021} is based on inverse reinforcement learning that makes use of the representation of an agent that moves around according to a set of actions and a learned policy.
The original paper considers  a chessboard-like road network, for which only the actions `forward', `left', and `right' are included. Generally, the set of actions needs to be specified and for each intersection, all options need to be defined. In chapter \ref{sec:datasets} we elaborate on our adaptions to allow for more heterogeneous road networks.
The model does not provide any privacy guarantees or mechanisms. However, we still include the model in our evaluation to elaborate on its suitability and the potential for further DP advancements.

\section{Utility measurement}
\label{sec:def-utility}

Defining and evaluating the utility of synthetic mobility data is a complex task. Unlike some other domains such as medical data, where typical applications like diagnosis prediction can be framed as classification tasks, typical mobility data applications do not lend themselves easily to such formulations.
As a result, utility is often assessed based on the degree of statistical similarity between the synthetic and raw data for different \textit{mobility characteristics}. 
A mobility characteristic, like the spatial distribution, is \textit{operationalized} by concrete measures, like the Jensen-Shannon divergence (JSD) of visits per grid cell.
Accordingly, hitherto executed evaluations of the considered synthesis models are based on similarity measures concerned with 
all or some of the following characteristics: spatial distribution, OD pairs, frequent patterns of sequences, and trip lengths.
However, the measured level of utility is highly dependent on the selected characteristics and their operationalization. 
For example, a high similarity of the spatial distribution does not necessarily ensure that the generated trips consist of reasonable OD combinations.
Furthermore, achieving high similarity based on, e.g., the JSD given a coarse grid with large cells (e.g., $2km$ x $2km$) provides only limited insight into the preservation of spatial distribution at a more detailed level.

Moreover, assessing utility based solely on these measures may not provide a comprehensive understanding of how well the synthetic data performs in real-life scenarios, like the examples given in the introduction. To overcome this limitation, we propose adopting a practitioner's perspective to specifying high utility for synthetic trip data and identifying appropriate measurements. By doing so, we aim to capture the utility of synthetic data more effectively in real-world applications.

Mobility data often encompasses more than just geo-locations, including temporal information, mobility modes, and user demographics. These additional aspects contribute to a broader range of characteristics that practitioners may find essential \cite{kapp_collection_2022}. However, none of the five evaluated models considers user-related information, and only one model generates timestamps. Consequently, only characteristics addressing solely geo-locations can be meaningfully evaluated, reducing the set of characteristics to the distribution of OD counts, trip lengths, and the spatial distribution of records.

Since the distribution of OD counts does not depend on the specifics of trip data, we decided to exclude respective evaluations and focus on \textit{trip lengths} and \textit{spatial distributions}, as they rather allow to capture the inherent complexity of fine-granular \textit{sequences} resembling street-level movements.

\subsubsection*{Map matching}
Tasks based on trip data are commonly based on a street-level granularity, such as assessing traffic volumes on roads or occupancy information of public transit lines.
However, all models under evaluation are based on grids that do not consider road networks and thus produce implausible trips that, e.g., cross buildings or rivers.
Thus, we introduce an additional processing step of map matching such that all synthetic trips are `snapped' onto the closest road, and consecutive records are connected according to the laws of physics and road regulations. We refer to data as produced by the model as \textit{original} and its matched variation as \textit{matched}.
In Section \ref{sec:matching}, we evaluate the suitability of considered synthetic datasets for map matching.

\subsubsection*{Routing as baseline}
Furthermore, for the respective comparisons, we introduce routing as a baseline.
Routing engines such as Google Maps, the Open Source Routing Machine (OSRM), or OpenTripPlanner use road and public transit network data to generate routes, typically optimized for the shortest or fastest paths. Thus, these engines offer an efficient and privacy-preserving approach to create realistic routes between a start and an endpoint. However, it's important to note that routing does not reflect real-life route choices comprehensively, as it cannot consider all factors relevant to individuals. For example, a cyclist might prefer a route through a park, even if it takes a bit longer, in order to avoid cobblestone streets or streets with too much car traffic. This highlights the value of actual user data for the work of practitioners  \cite{lu_understanding_2018, broach_where_2012, prato_route_2009}. We argue, that synthetic data generation algorithms need to outperform routing engines 
to provide justifiable additional value. 
A \textit{routed} variation of a dataset is created by routing the identical start and destination location of each included trip.

\subsubsection*{Trip lengths}
Existing assessments of trip lengths generally indicate that generated synthetic data reaches a satisfactory level of similarity to the raw data.
However, these evaluations only consider implausible unmatched trips, thus  the trip length
says little about the actual traveled distance if one were to follow the route. We argue that  utility concerning trip length can only be considered high when taking into account the trip length of the route matched to the road network. 
To identify synthetic trips that  are implausible due to  unreasonable loops and unnecessary winding paths, we suggest conducting a secondary evaluation of trip length which is based on comparing the synthetic trips to the straight-line (SL) distance between their origin and destination. 

\subsubsection*{Spatial distribution}
To capture the spatial distribution of trips on a street level, a sufficiently fine-grained grid is required for evaluations.
As a rough guideline, we consider grid cells with dimensions of $1 km$ x $1 km$ to capture a neighborhood 
and cells of $40 m$ x $40 m$ provide details on the level of roads.
As a proxy for \textit{traffic volumes} per road segment, we suggest evaluating the spatially aggregated trips based on a grid with a resolution on street level (of $40m$) by computing the Jensen-Shannon-divergence (JSD) between the matched raw dataset and the variants of synthetic datasets, where each cell is considered as a `segment'. 

\begin{figure}[tb]
    \centering
    \includegraphics[width=0.4\textwidth]{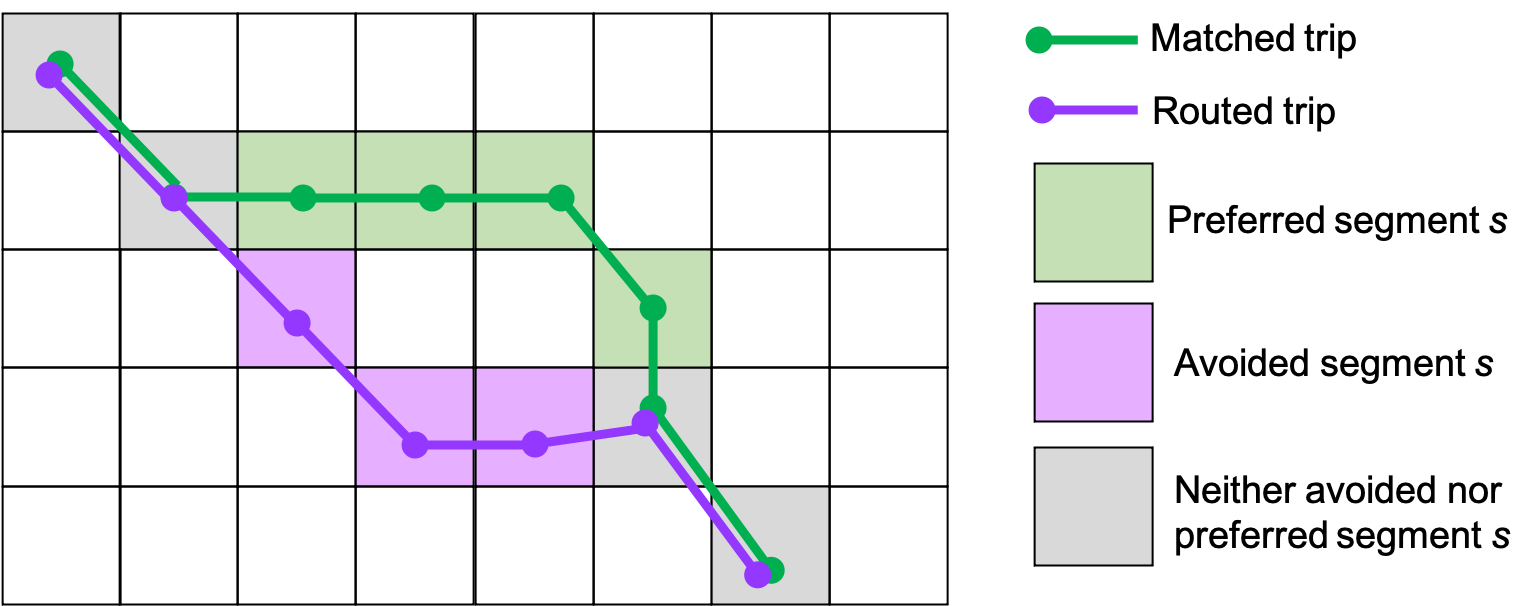}
    \caption{Schematic illustration: preferred segments (green cells) are traversed by the matched trip (green line), but not its routed counterpart (purple line), and vice versa, they are considered avoided (purple cells).}
    \label{fig:preferenceExample}
\end{figure}

It is difficult to interpret the values of the JSD  in terms of practical utility. We thus additionally propose the downstream task of identifying avoided and preferred roads.
For each trip, the matched and routed versions are compared; see Fig. \ref{fig:preferenceExample} for a visual example. Each road segment $s$ that is traversed by the matched but not the routed trip is considered `preferred' and the counter $\mathrm{npref}_{s}$ is increased by 1. Conversely, each segment that is traversed by the routed but not the matched trip is considered `avoided' and its counter $\mathrm{navoid}_{s}$ is increased by 1. 
We define the preference score of $s$ as  
$$\mathrm{preferenceScore}_s = \frac{\mathrm{npref}_s -\mathrm{navoid}_s}{n_s}, $$
where $n_s$ equals the total number of times that $s$ has been passed by a matched or a  routed trip. 

Preference scores can be assessed statistically, e.g., using the Pearson correlation coefficient between scores from raw and synthetic data. Secondly, the correct classification (preferred > 0, avoided < 0, neither = 0) can be analyzed in terms of standard classification metrics such as accuracy and F1-scores.
To approach the downstream task from a practical perspective, we suggest a measurement of human inference, such as a survey. See Section \ref{sec:survey} for our realization.

Finally, we propose to assess the spatial distribution with respect to traffic flow at intersections which is another relevant application \cite{li_deriving_2010} 
for example for traffic signal management \cite{wei_survey_2020}. 
In this context, we refer to traffic flow as the entirety of movement patterns at an intersection. A movement pattern is described by the arrival road segment, the turning behavior, and the destination road segment. Similar patterns are grouped into movement clusters.
As trips are already matched to the road network, movement patterns can easily be aggregated by clustering trips spatially based on a suitable distance metric. We refer to Section \ref{sec:clustering} for our utilized implementation.

In summary, we propose to measure the utility of trip data by comparing synthetic data, after it has been matched with the road network, to its routed counterpart with respect to the following aspects: (1a) distribution of trip lengths, (1b) distribution of trip lengths in relation to the OD straight-line distance, (2) distribution of traffic volumes on road segments, (3) identification of avoided and preferred roads, and (4) traffic flow at large intersections.

\section{Experimental setup}
\label{sec:exSetup}

\subsection{Raw dataset}
\label{sec:datasets}

For our evaluation, we use the openly available SimRa dataset \cite{bermbach_simra_2021, bermbach_simra_2021-1, karakaya_simra_2022}. SimRa \cite{karakaya_simra_2022} is a smartphone app for cyclists that records GPS traces, sampled every few seconds, to track near-miss incidents.
 We use the trips collected in the central area of Berlin limited to records within longitudes of 13.267 and 13.482, and latitudes between 52.445 and 52.576, thus covering about 14x14 $km^2$. The dataset comprises 28,345 trips between 2019 and 2022. Near-miss incident information is discarded for our evaluation.

\subsection{Configurations for synthesis models}

For each evaluated synthesis algorithm, including both the differentially private (DP) and non-DP variants, when applicable, we generate a synthetic dataset containing 5,000 trips. To account for the non-deterministic nature of model-based data generation, we repeat this procedure five times and average the results. A privacy budget of $\varepsilon=2$ is used for the DP condition, which is the highest privacy budget setting in the evaluations of AdaTrace and PrivTrace. As the implementations of AdaTrace and PrivTrace do not provide a non-DP option, we set a very large privacy budget of $\varepsilon=1,000,000$ for the base condition `without DP'. Datasets are generated on a machine equipped with an Nvidia A30 GPU, an Intel(R) Xeon(R) Gold 6346 CPU, and 251 GB RAM.

The choice of spatial granularity is pivotal to our utility formulation. Thus, we employ at the very least the finest granularity proposed by the respective authors. When computationally feasible, we opt for even finer resolutions.
It is noteworthy that AdaTrace uses an adaptive grid, adjusting the spatial granularity based on the density within each grid cell. However, as explained in Section \ref{sec:algorithms}, the distance between consecutive points is determined by the top-level grid, which defaults to 6 x 6 cells in the respective source code. For the utilized SimRa dataset this would result in a top-level cell size of $2.5 km$ x $2.5 km$ and an average sequence length of 3.6 points. 
Such distances between consecutive points would not meet the requirements with respect to our utility measurement, as stated in Section \ref{sec:def-utility}. We thus adjust the top-level grid to $N=28$ which corresponds to a cell size of $500 m$ x $500 m$. A finer-grained grid with a cell size of $250 m$ x $250 m$ exceeds the available memory.

The spatial granularity of the top-level grid of PrivTrace is determined by a parameter $K$.
We use a resolution of $500m$ ($K=28$) for the sake of computational feasibility. Using a finer grid with a resolution of $250m$ significantly increases the runtime from 15 minutes to 34.5 hours.
We further adjust AdaTrace and PrivTrace to generate coordinates with a precision of four decimals, as both round to only two decimals, which would correspond to a granularity of  $\sim 1 km$.

The BiLSTM uses a similar neural network architecture as \cite{blanco-justicia_generation_2022}. To speed up training and prevent overfitting, we implement an early stopping criterion causing the training to stop if the validation loss increases or stagnates for more than one epoch.
Also, to enhance utility, we draw start locations from the distribution of start locations in the raw data instead of relying on random sampling. 
We use a grid resolution of $250m$, which is roughly equivalent to the authors' fine-granular choice. Additionally, we tested a finer resolution of $100m$, however, we discarded it due to computational infeasibility as it takes 8 hours for training and about 25 hours for trip generation, compared to 75 minutes for training, and 8 hours for generating when a resolution of $250m$ is employed.
 Like the authors, we modify the final dense layer based on the 3,188 cells that have been visited at least once.
We also conduct a preprocessing step where all trips above a given trip length are discarded to save training resources and set the maximum length to 50 steps.
In our setup, we yield an early stopping after 16 epochs, a training loss (accuracy) of 0.44 (86.67\%), and a validation loss (accuracy) of 0.54 (85.68\%). 

To ensure DP, we apply the Laplace mechanism to the distribution of start locations and trip sequence lengths, respectively. 
As a fine-granular grid is expected to decrease utility in a DP setup due to a higher level of required noise, we initially test the DP variation with a coarser resolution of $500m$. The BiLSTM final dense layer is accordingly  adjusted to 900 cells.
The BiLSTM utilizes the DP Adam optimizer with an L2 clipping norm of 1, a noise multiplier of 1.4, and 64 micro-batches. 
The privacy budget is split such that $\varepsilon=1$ is allocated for the BiLSTM and $\varepsilon=0.5$ for each distribution. 
The batch size is reduced to 128 (due to memory limits), and 20 epochs are trained without early stopping. It takes 46 hours of training, thus a multitude of the non-DP version, yielding a training loss (accuracy) of 5.59 (10.56\%) and a validation loss (accuracy) of 5.41 (12.82\%). Due to unusable results for map matching, as we will show in the following Section \ref{sec:matching}, we refrain from evaluations with a finer grid resolution.

We run DP-Loc with a $500 m$ grid resolution.
As the output produces multiple datasets for different iterations of a Metropolis-Hastings algorithm, we follow the authors' advice for smaller datasets and use 100 iterations. Again, due to results unsuitable for map matching (see Section \ref{sec:matching}) we do not test further resolutions.

To employ TrajGAIL without translating the Berlin road network to the TrajGAIL network notation, we define the network based on a grid. For each cell, its eight neighboring cells are considered to be within reach according to one of eight respective actions. We project the dataset to a grid with a $500m$ resolution which results in a significantly larger network compared to the dataset evaluated by the authors of TrajGAIL, which only encompasses 9 intersections, and thus substantially higher computation times. While one iteration for their dataset only takes 8 seconds, the processing time for the SimRa dataset extends to approximately 3.5 minutes. Consequently, running 20,000 iterations, similar to the authors' evaluation, would require over 48 days to complete. Due to this computational infeasibility, TrajGAIL is excluded from further evaluations.

\subsection{Matching and routing implementation}

\begin{figure}[tb]
    \centering
        \includegraphics[width=0.23\textwidth]{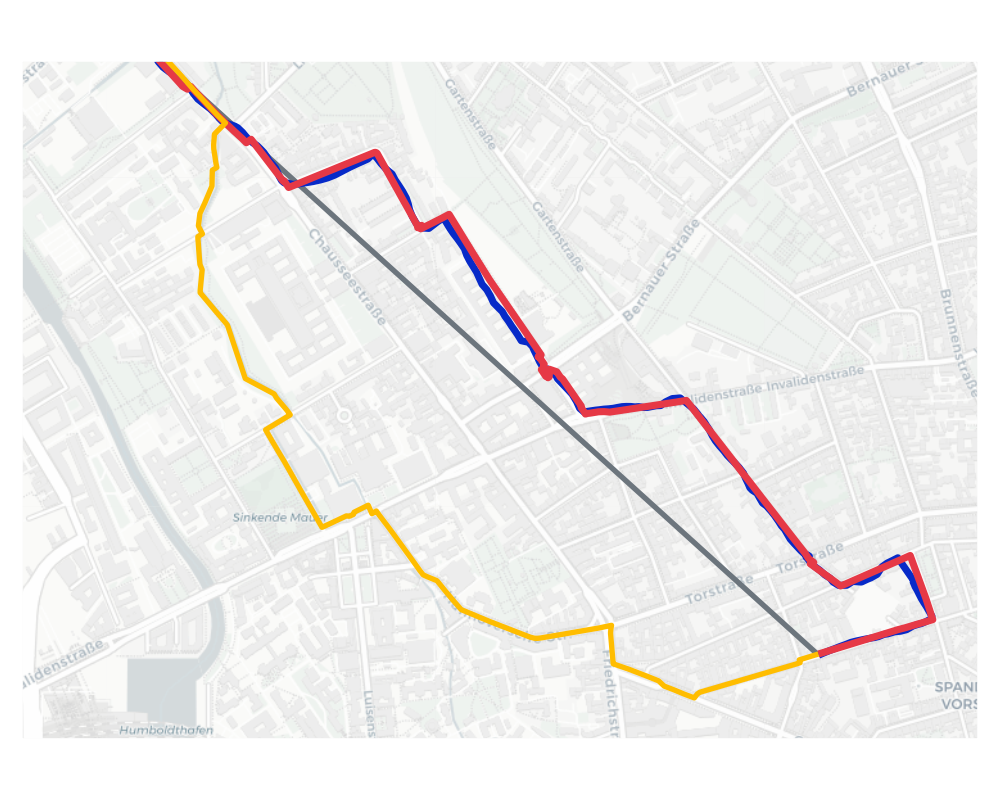}
        \includegraphics[width=0.23\textwidth]{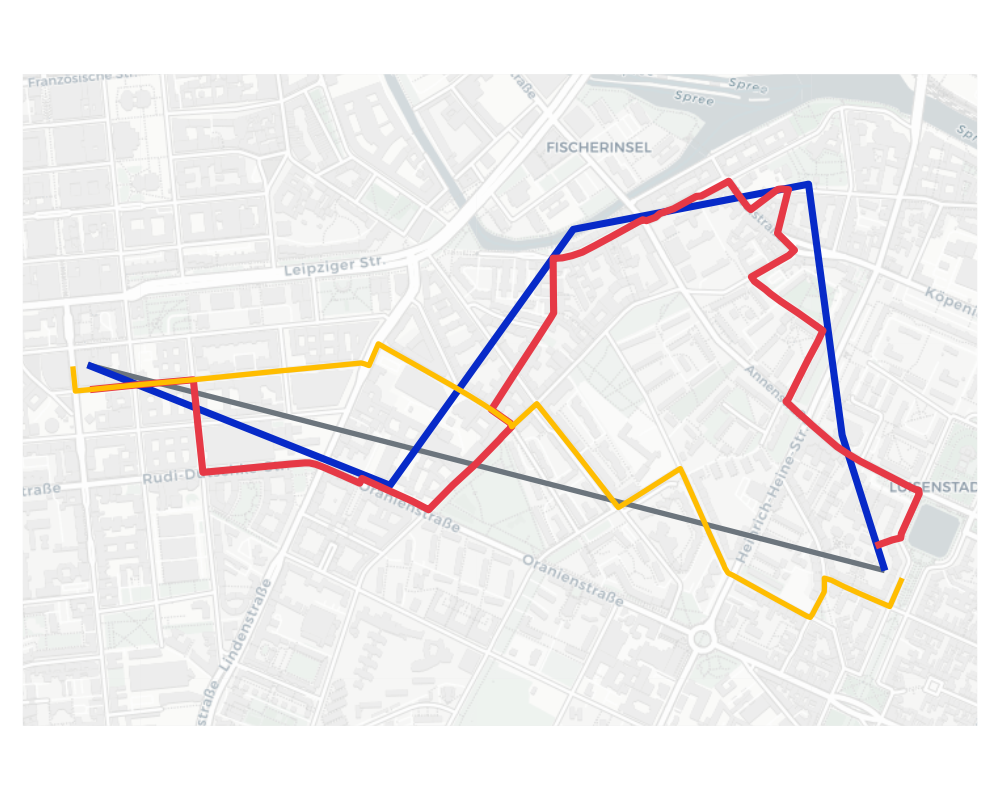}
        \begin{tikzpicture}
            \node[font=\footnotesize] {Basemap: OpenStreetMap contributors};
        \end{tikzpicture}
        \input{figures/matching_legend}
        \vspace*{-4cm}
    \caption{Example for raw (left) and AdaTrace (right) trip in four variations.}
    \label{fig:map_matching}
\end{figure}
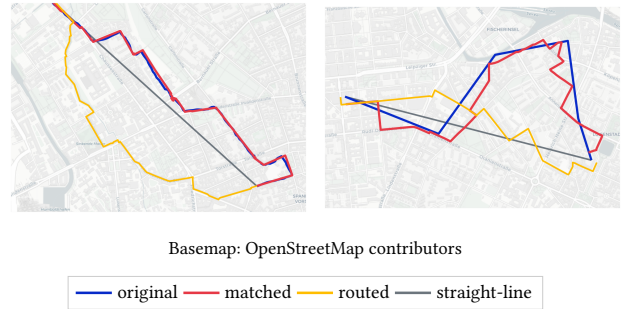

We use the OpenStreetMap Routing Machine (OSRM) \cite{luxen-vetter-2011} \textit{match service (API)} to snap trips onto the OpenStreetMap road network. 
The parameter $radius$ of the API is intended to set the GPS precision. We chose a setting of 10 meters for the raw and 20 meters for synthetic datasets which proved to be a reasonable setting within acceptable computation time.\footnote{Note that map matching is computationally expensive;  usage in practice would require optimization such as parallelization to be feasible.}
To set up OSRM, we used the OpenStreetMap data excerpt for Berlin, provided as $PBF$ file by Geofabrik\footnote{https://www.geofabrik.de/data/download.html}, from 17th November 2022 using the bicycle traffic mode. 
There are cases in which OSRM map matching does not produce entirely accurate results; see Appendix \ref{appendix:matching}  for potential issues and respective validation analyses.
For the routing baseline, we use the OSRM \textit{Route Service} based on a similar OSRM setup. 
See Figure \ref{fig:map_matching} for an example of created trip variations.

\subsection{Road preference survey}
\label{sec:survey}

\begin{figure}[tb]
    \centering
    \includegraphics[width=0.23\textwidth]{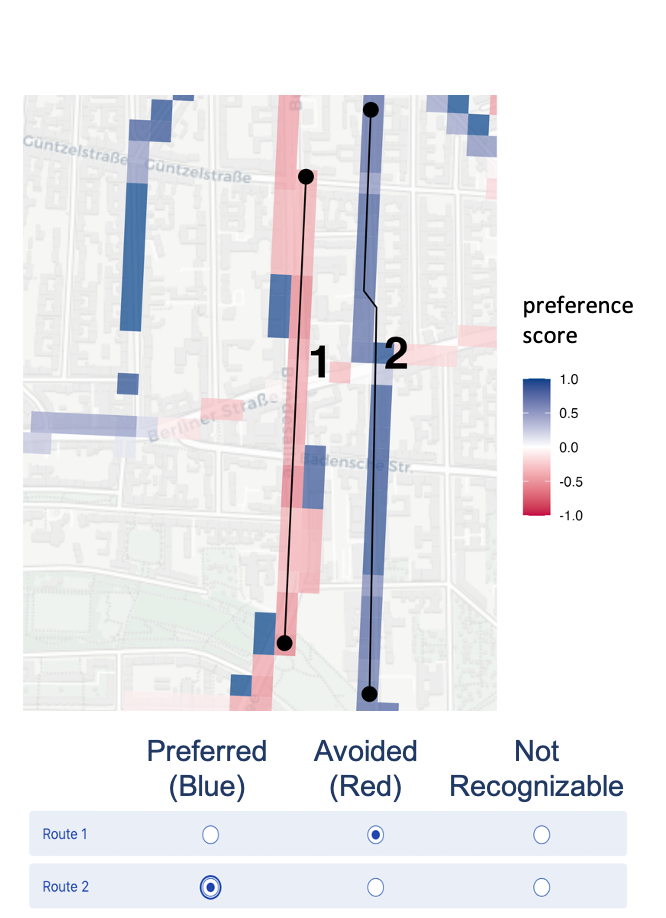}
    \hfill
    \includegraphics[width=0.23\textwidth]{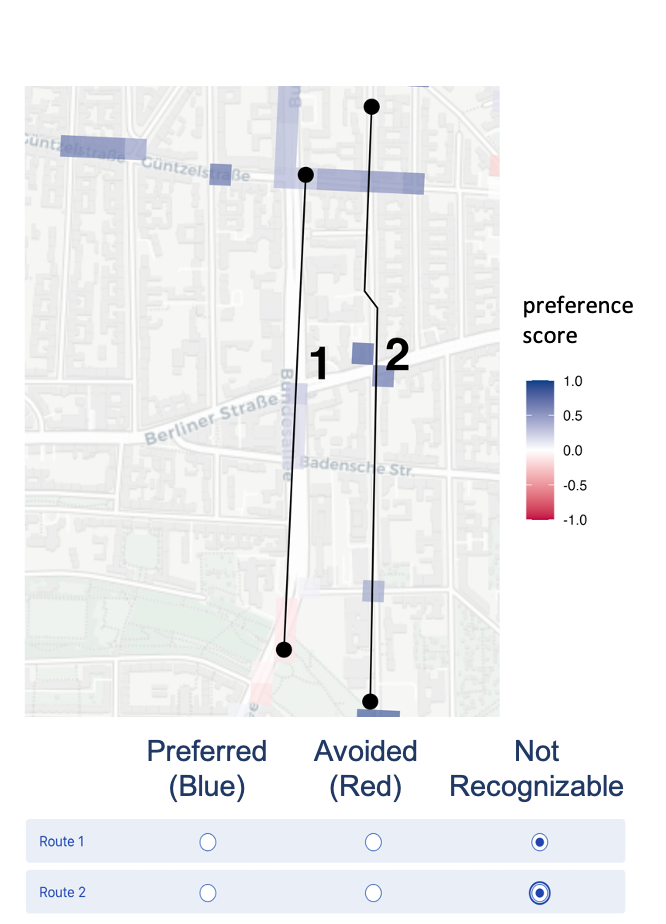}
    \caption{Examples of survey questions. The preference score is overlaid on a map cutout alongside selected roads. Answers are given via radio buttons for each road.}
    \label{fig:surveyExample}
\end{figure}

We evaluate the downstream task of road preference detection with a survey on 40 selected meaningful examples of roads.
Participants are asked to decide for each road, whether they consider it to be `avoided', `preferred', or `not recognizable', based on the displayed preference score. To select roads, we initially filter segments that have clearly either been preferred or avoided by restricting to those with a preference score $> 0.5$ or $< -0.5$. 
Next, we limit the selection to the top 10\% of visited and routed segments to focus on highly frequented areas.
Finally, to obtain coherent clusters representing entire roads that are either avoided or preferred, adjacent remaining segments of the same class are merged. 20 avoided and 20 preferred roads are selected from the largest clusters.
The survey consists of map cutouts displaying preference scores alongside selected roads (see Figure \ref{fig:surveyExample}). To enhance clarity, only preference scores from the top 25\% of frequented segments are shown which are more likely to be robust and representative.
The survey is conducted online, in German, and for each condition, at least 10 people are recruited. Each questionnaire contains only one condition to prevent learning effects.

\subsection{Traffic flow evaluation}
\label{sec:clustering}

We evaluate traffic flow preservation based on ten major intersections. They are selected by initially filtering the top 15 most visited cells from the raw data based on a grid with $500m$ resolution and manually setting ten meaningful bounding boxes within them.
For each intersection and each routed and matched dataset, trips are cut according to the respective bounding box. The pairwise distance between these trips is computed with the Hausdorff distance (HD). Complete-linkage agglomerative hierarchical clustering is employed based on the resulting distance matrix. To determine movement clusters, the hierarchical tree is cut at a height of 5 meters \footnote{In experiments using various cutoff values, the relatively strict value of 5 proved to be the most suitable, as it effectively prevents false positives while minimizing false negatives (since trips are matched to the same road network, resulting in minimal variance).}

To assess the preservation of movement patterns by synthesis algorithms, the clusters of each synthetic dataset are linked with the clusters of the raw data. 
To link clusters, we employ a simple greedy matching algorithm that iterates over all clusters in the raw data. For each raw data cluster, we find the synthetic data cluster with the smallest HD. If the distance is smaller or equal to 5, they are considered a match, otherwise, the raw data cluster remains unmatched. 
To ensure computational feasibility, not all trips, but randomly selected representatives from each cluster of both datasets are used for distance computation. 
Finally, we compare the counts for each movement pattern using the normalized discounted cumulative gain (nDCG) which is a common measure to assess ranking quality in information retrieval. Only the $k$ highest ranked scores are considered. The normalized version has a range from 0 to 1, where a score of 1 signifies perfect retrieval.

\section{Results}
\label{sec:results}

\subsection{Matching ability}
\label{sec:matching}
The longer the distance between consecutive points the less suitable is the application of map matching; in the extreme case where a trip would only include origin and destination points, map matching would be equal to routing. Thus, for synthetic data to be usable for map matching we require the distance between consecutive points for the majority of trips to be within the magnitude of the spatial resolution used for data generation.\footnote{Note, that the raw data also includes outliers of jumps up to $10 km$ where some are due to faulty GPS sensor signals and others because a trip leaves and re-enters the defined bounding box at a different location.}
Table \ref{tab:consecDist} shows the distribution of distances between two consecutive points for each synthetic dataset (excluding repetition of identical locations).

\begin{table}[tb]
\caption{Distribution of distances between two consecutive points (excluding repetition of identical locations) in meters.}
\begin{tabular}{lrrrrr}
\toprule
{} &  min &   25\% &  median &   75\% &    max \\
\midrule
raw           &    1 &    11 &      16 &    20 &  10,197 \\
\midrule
AdaTrace      &    7 &   441 &     589 &   746 &   1,643 \\
AdaTrace DP  &    7 &   501 &     664 &   835 &   2,008 \\
PrivTrace     &    2 &   374 &     526 &   677 &   1,135 \\
PrivTrace DP &   15 &   329 &     486 &   649 &   1,109 \\
BiLSTM           &  248 &   249 &     250 &   250 &  18,513 \\
BiLSTM DP       &  498 &  2,999 &    5,226 &  8,044 &  17,338 \\
DP-Loc         &  498 &   999 &    1,412 &  2,554 &  13,436 \\
Dp-Loc DP     &  498 &   999 &    1,500 &  4,607 &  14,304 \\
\bottomrule
\end{tabular}
\label{tab:consecDist}
\end{table}

 While AdaTrace (DP), PrivTrace (DP), and BiLSTM yield distances in the required order of magnitude, DP-Loc proves not suitable for map matching with a median distance of $1.4 km$ ($1.5 km$ for the DP version). An examination of the trips generated by DP-Loc revealed that after only a few initial points a large jump to the destination followed. As our evaluation is based on road network matching, we discard DP-Loc from further analyses as well as the DP version of BiLSTM, which only consists of random jumps with a median distance of $5 km$. This yields the following final set of synthetic datasets for the evaluation: \textbf{AdaTrace}, \textbf{AdaTrace DP},  \textbf{PrivTrace}, \textbf{PrivTrace DP}, and \textbf{BiLSTM}.

See Appendix \ref{appendix:vizOriginal} for visualizations of spatial distributions and example trips of each original dataset (without DP) and Appendix \ref{appendix:spatialAgg} for spatial distributions of matched and routed variations.

\subsection{Trip lengths}
\label{sec:tripLenghRes}

As shown in Table \ref{tab:tripLength},  
all synthetic median SL distances tend to be shorter than the raw data's SL distance of $4.57 km$.
AdaTrace is most similar and within a reasonable distance, followed by BiLSTM,  while the other three synthetic data are too far off. More precisely, PrivTrace is much shorter with a median distance of $1 km$ and even only $470 m$ for its DP version, while the SL distance of AdaTrace DP is a lot longer than that of the raw data.
In contrast to the described tendency for SL distances, the original trips of AdaTrace and BiLSTM tend to be longer than the original raw trips. Interestingly, despite AdaTrace DP having the furthest deviation in terms of absolute value, its SL ratio is closest to that of the raw data.
While PrivTrace (DP) remains significantly below the respective value of the original data, the increase compared to the straight line is the highest with a median of 216\% (184\%) of the SL distance. 
These observations suggest that synthetic data is more circuitous than real-life trips.
This effect is even more evident in the matched datasets, as the SL ratio increases even more across all models. 

\begin{table}[tb]
\caption{Median of trip lengths in $km$ for each dataset and their variations. In brackets: median ratio of respective trip length to its straight-line (SL) distance counterpart.}
\begin{tabular}{lrrrrrrr}
\toprule
 & SL & \multicolumn{2}{c}{original} & \multicolumn{2}{c}{matched} & \multicolumn{2}{c}{routed} \\
\midrule
raw  &  4.57 &  5.79 &  (120\%) &    5.47 &  (130\%) &   5.63 &  (120\%) \\
\midrule
AdaTr. &   4.10 &  7.45 &  (170\%) &    9.01 &  (212\%) &   5.12 &  (120\%) \\
Ada. DP  &  6.84 &  8.43 &  (120\%) &    9.71 &  (150\%) &   8.47 &  (120\%) \\
PrivTr.  &  1.02 &  2.55 &  (216\%) &    3.15 &  (268\%) &   1.42 &  (130\%) \\
Priv. DP  &  0.47 &  1.05 &  (184\%) &    1.28 &  (229\%) &   0.73 &  (140\%) \\
BiLSTM     &  3.75 &  7.31 &  (162\%) &     7.89 &  (200\%) &   4.65 &  (120\%) \\
\bottomrule
\end{tabular}
\label{tab:tripLength}
\end{table}

As expected, the routing algorithm reliably produces trips with a median ratio to the SL distance of 120\%-140\% for all models. In comparison to the matched synthetic trips, the routed lengths are thus much closer to those of the raw trips 
and, for AdaTrace (DP) and BiLSTM, also with regards to the absolute value. 
Thus, none of the models succeeds at producing plausible trip lengths.

\subsection{Traffic volume}

Table \ref{tab:roadSegUsage} shows the JSD between matched raw data and all dataset variations, based on aggregations on a grid with a $40m$ resolution\footnote{Note that we do not only account for cells that contain a record but also those that are passed by the line from one record to the consecutive.}.
The JSD of the synthetic data SL versions are all in the same order of magnitude as the JSD of the raw SL version, providing a baseline upper bound of the JSD for each synthetic dataset. (Recall that the higher the value the lower the similarity of the respective datasets.) The spatial distribution of AdaTrace and PrivTrace original datasets, with and without DP, improve compared to the straight-line distribution, though only marginally. The JSD for the original BiLSTM dataset even increases.

\begin{table}[tb]
\caption{JSD values for traffic volumes between each dataset variant and the baseline of matched raw data. Bold font indicates the best variation. }
\begin{tabular}{lrrrr}
\toprule
   &  SL &  original &  matched &  routed \\
\midrule
raw &  0.60 &      0.15 &     0.00 &    0.39 \\
\midrule
AdaTrace &  0.60 &      0.58 &     \textbf{0.39} &    0.40 \\
AdaTrace DP &  0.61 &      0.60 &     \textbf{0.43} &    0.46 \\
PrivTrace  &  0.63 &      0.59 &     \textbf{0.43} &    0.45 \\
PrivTrace DP &  0.69 &      0.67 &     \textbf{0.58} &    0.60 \\
BiLSTM    &  0.59 &      0.62 &     0.44 &   \textbf{0.40} \\
\bottomrule
\end{tabular}
\label{tab:roadSegUsage}
\end{table}

All JSD values for the matched versions show the expected decrease, supporting our claim that map matching is needed for usable results on a street level. For AdaTrace and PrivTrace the results outperform their routed counterpart, at least slightly. On the contrary, the routed version of BiLSTM outperforms its matched one.
The better performance of AdaTrace and PrivTrace comes likely due to their adaptive grid which captures `hotspots' on a more fine granular level than the top-level grid. The low  performance of the BiLSTM model is presumably rooted in the static grid utilized for training and data  generation, and the difference between the grid resolution chosen in training ($250 m$) vs. evaluation ($40 m$) that reflect street-level relations. 
Experiments show that a JSD based on a coarser grid using a $250 m$ resolution yields a superior value also for the matched BiLSTM data in comparison to the routed version. 
Thus, the BiLSTM might only obtain superior results for its matched data on a street level if a more fine-granular resolution is used for data generation.
It is thus interesting to note that the map matching is not sufficient to compensate for the coarse resolution used for data generation.

AdaTrace shows the best overall results. It yields the only competitive JSD with its matched version to the routed raw version. Also, the JSD values of AdaTrace DP are only slightly worse than AdaTrace without DP. Unlike PrivTrace, which shows a majorly decreased performance for its DP version.

\subsection{Road preference}

\subsubsection{Statistical similarity}

We assess the preservation of road preference in synthetic data by comparing their preference scores with the ground truth obtained from the raw data. 
We test variations of two hyperparameters for the classification task. (1)  Rarely visited segments are discarded based on a \textit{frequency threshold}. 
Different thresholds are evaluated: top-100\%, top-75\%, and top-25\% of frequented segments, based on the series of all $n_s$, are considered. 
Note that a threshold of top-100\% frequented segments means that all segments with at least 1 record are included.
(2) To account for cases where values are close but do not fall into the same class (e.g., the preference score for a segment is 0.1 according to the raw and -0.1 according to the synthetic dataset), we set a \textit{tolerance level} such that two values are considered correctly classified if they do not differ more than the tolerance level. We consider values of 0 and 0.3.
As shown in Table \ref{tab:preferenceScore}, the BiLSTM clearly outperforms AdaTrace and PrivTrace on all settings with respect to the correlation coefficient, the accuracy, and the F1-score of the class `avoided', while it is outperformed in terms of the F1-score of the `preferred' class in three out of four hyperparameter settings though with less pronounced differences. As expected, setting a tolerance level increases the scores, as classes are assigned more generously. 
Further, intuitively almost all scores increase with an increasing frequency threshold, as highly frequented cells are expected to be captured with higher accuracy by the models.

\begin{table}[tb]
    \centering
\caption{Preference score evaluation for different frequency thresholds (as top-$x$\% frequented segments) and tolerance levels based on Pearson correlation coefficient $r$, accuracy and F1-score, precision, and recall for the classes `avoided' and `preferred'. For each hyperparameter setting, i.e., each $n$-th row per block, the best value is depicted in bold font.}
\begin{tabular}{p{0.6cm} p{0.5cm}p{0.4cm}|p{0.3cm}p{0.4cm}p{1.8cm}p{1.8cm}}
\toprule
 & frequ. thres. & tol. lev. &  \centering $r$ &  \centering accu. (\%) & \centering avoided  F1~[prec.,~rec.] &  preferred  F1~[prec.,~rec.] \\
\midrule
Ada              & 100 & .0 &         .40 &           56 &          .46 [.64, .36] &             \textbf{.70} [.58, .89] \\
Trace            & 75  & .0 &         .46 &           58 &          .48 [.69, .37] &             \textbf{.72} [.60, .90] \\
                 &     & .3 &         .46 &           63 &          .53 [.77, .41] &             \textbf{.74} [.62, .92] \\
                 & 25  & .3 &         .57 &           68 &          .64 [.82, .52] &                      .75 [.65, .90] \\
\midrule
Ada     & 100 & .0 &         .37 &           53 &          .51 [.54, .48] &             .65 [.55, .80] \\
Trace   & 75  & .0 &         .40 &           55 &          .53 [.57, .49] &             .67 [.58, .80] \\
DP&           & .3 &         .40 &           60 &          .57 [.63, .52] &             .69 [.60, .83] \\
        & 25  & .3 &         .44 &           66 &          .70 [.70, .70] &             .66 [.65, .67] \\
\midrule
Priv             & 100 & .0 &         .27 &           51 &          .28 [.51, .19] &            .67 [.55, .85] \\
Trace            & 75  & .0 &         .31 &           53 &          .29 [.54, .20] &            .68 [.56, .88] \\
                 &     & .3 &         .31 &           57 &          .35 [.63, .24] &            .70 [.58, .90] \\
                 & 25  & .3 &         .41 &           59 &          .36 [.83, .23] &            .71 [.57, .94] \\
\midrule
Priv     & 100 & .0 &         .19 &           48 &          .33 [.43, .26] &              .64 [.59, .69] \\
Trace    & 75  & .0 &         .19 &           48 &          .33 [.43, .26] &              .64 [.59, .69] \\
DP&            & .3 &         .19 &           55 &          .40 [.52, .33] &              .67 [.62, .73] \\
         & 25  & .3 &         .23 &           59 &          .43 [.62, .32] &              .70 [.60, .83] \\
\midrule
Bi      & 100 & .0 &         \textbf{.43} &           \textbf{58} &          \textbf{.59} [.59, .60] &            .67 [.63, .72] \\
LSTM    & 75  & .0 &         \textbf{.48} &           \textbf{61} &          \textbf{.62} [.63, .62] &            .69 [.64, .76] \\
        &     & .3 &         \textbf{.48} &           \textbf{66} &          \textbf{.66} [.67, .66] &            .72 [.67, .78] \\
        & 25  & .3 &         \textbf{.59} &           \textbf{72} &          \textbf{.73} [.81, .66] &            \textbf{.77} [.69, .87] \\
\bottomrule
\end{tabular}
    \label{tab:preferenceScore}
\end{table}

The class `preferred' is detected best by all models.
This is especially true for AdaTrace (DP) and PrivTrace (DP) and can likely be explained with the selection of coordinates in their generation algorithm: Recall that data generation is based on discrete grid cells. While the BiLSTM always uses the center of the grid cells to create coordinates, AdaTrace and PrivTrace sample a random point from within the cell. This produces more variation and thus more visited cells (see Appendix \ref{appendix:spatialAgg} for maps of respective spatial distributions). Therefore, there is a bias towards `preferred' classification: AdaTrace (PrivTrace) classifies 79\% (79\%) of considered cells as preferred and only 17\% (13\%) as avoided, while only 67\% of cells are preferred in the raw data and 36\% avoided. Thus, the high `preferred' F1-score is especially driven by a high recall. 
In summary, the BiLSTM can be considered the best-performing model in terms of road preference and avoidance. 

Considering DP models for both AdaTrace and PrivTrace, there is only a slight decrease in terms of accuracy, while the correlation coefficient decreases to a larger extent. 
Interestingly, both PrivTrace DP and Adatrace DP detect avoided roads slightly better than their non-DP counterparts. Further investigation showed that this is a noise-induced artifact causing more variation in OD pairs and thus a higher share of routed cells (24\% for both, AdaTrace and PrivTrace) which leads to an increased recall of `avoided' and a decrease for `preferred', while the precision either decreases or remains the same for all  classes.

The outcomes do not indicate a perfect classification but suggest that models have the potential to identify preferred and avoided roads to a certain extent, particularly when focusing solely on heavily trafficked sections.

\subsubsection{Survey}

We assess the homogeneity of answers per condition with Krippendorff's alpha $\alpha_K$ (see Table \ref{tab:percPrefClassification}) which is high for raw data and AdaTrace, while only medium for all other datasets ($\alpha_K \geq 0.66$). 
However, these discrepancies are seldom because participants directly contradict each other's assessments. Instead, they typically arise from certain participants adopting a more conservative stance, opting for `not recognizable' rather than making a definitive choice between `avoided' or `preferred'.
In particular, the results from the majority vote are robust when compared with sums of (in)correct answers.
Table \ref{tab:percPrefClassification} presents the percentage of accurately classified, incorrectly classified, and unrecognizable roads among the 40 selected roads, based on the majority vote of participants.

\begin{table}[tb]
\caption{Results of the road preference survey for 40 selected roads, presented as percentages of accurately classified roads in total and separated by class (preferred, avoided), the percentages of unrecognizable and incorrectly classified roads, and group homogeneity (Krippendorff's $\alpha_K$).}
\label{tab:percPrefClassification}
\begin{tabular}{lcccc}
\toprule
 &  $\alpha_K$ & accu. [pref., avoid.] &  not recog. &  incorrect\\
\midrule
raw        &  0.91  &        100   [100, 100] &  0 &    0 \\
\midrule
AdaTrace    & 0.80   &         82   [90, 75] &  15 &    2 \\
Ada. (DP)   & 0.74   &         75    [80, 70] &  20 &    5 \\
Priv.       & 0.66   &         52    [70, 35] &  42 &    5 \\
Priv. (DP)  & 0.66   &         30    [45, 15] &  68 &    2 \\
BiLSTM     & 0.73    &         80    [75, 85] &  20 &    0 \\
\bottomrule
\end{tabular}
\end{table}

An accuracy of $100 \%$ in raw routes supports our defined ground truth. 
Even for all synthetic datasets, 
misclassification of roads was minimal, with most discrepancies attributed to being classified as 'not recognizable'. This trend was particularly noticeable in PrivTrace (DP).
Generally, preferred roads are more often classified correctly than avoided ones, except for BiLSTM.
AdaTrace demonstrates the highest performance, closely followed by BiLSTM. Even in its DP version, 75\% of the roads are still correctly classified.
However, this also suggests that about every fourth road is misclassified or not recognizable. Considering that we only included the top 10\% most frequented cells for the selection of roads, these results presumably do not  suffice in practice.

Overall, the assessment of road preference indicates that especially AdaTrace and BiLSTM demonstrate a certain level of success in preserving spatial distribution information. We also find a set of inconclusive results with regard to different analyses concerned with spatial distribution: 
With regards to PrivTrace, even though its JSD slightly outperforms the BiLSTM's, it performs much worse in road preference classification. This is mainly driven by a bad detection of avoided streets and likely due to implausible short-distanced OD pairs, as seen in Section \ref{sec:tripLenghRes}.
Also, even though the BiLSTM obtains a slightly better F1-score for preferred segments than AdaTrace (DP) for the comparable hyperparameter setting of a top-25\% frequency threshold, the majority of individuals could identify 15 percentage points (5 percentage points) more of preferred roads based on AdaTrace (DP) data than on BiLSTM data.

\subsection{Traffic flow at intersections}

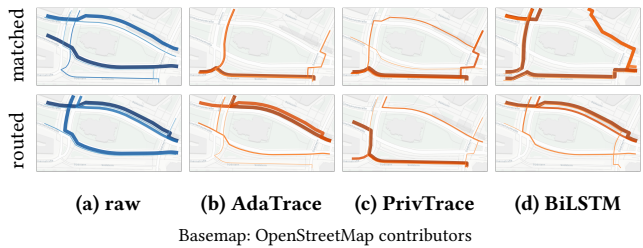
\begin{figure}[tb]
    \centering
    \begin{minipage}[t]{0.015\textwidth}
        \rotatebox[origin=l]{90}{\small matched}
    \end{minipage}
     \begin{minipage}[t]{0.11\textwidth}
        \includegraphics[width=\textwidth]{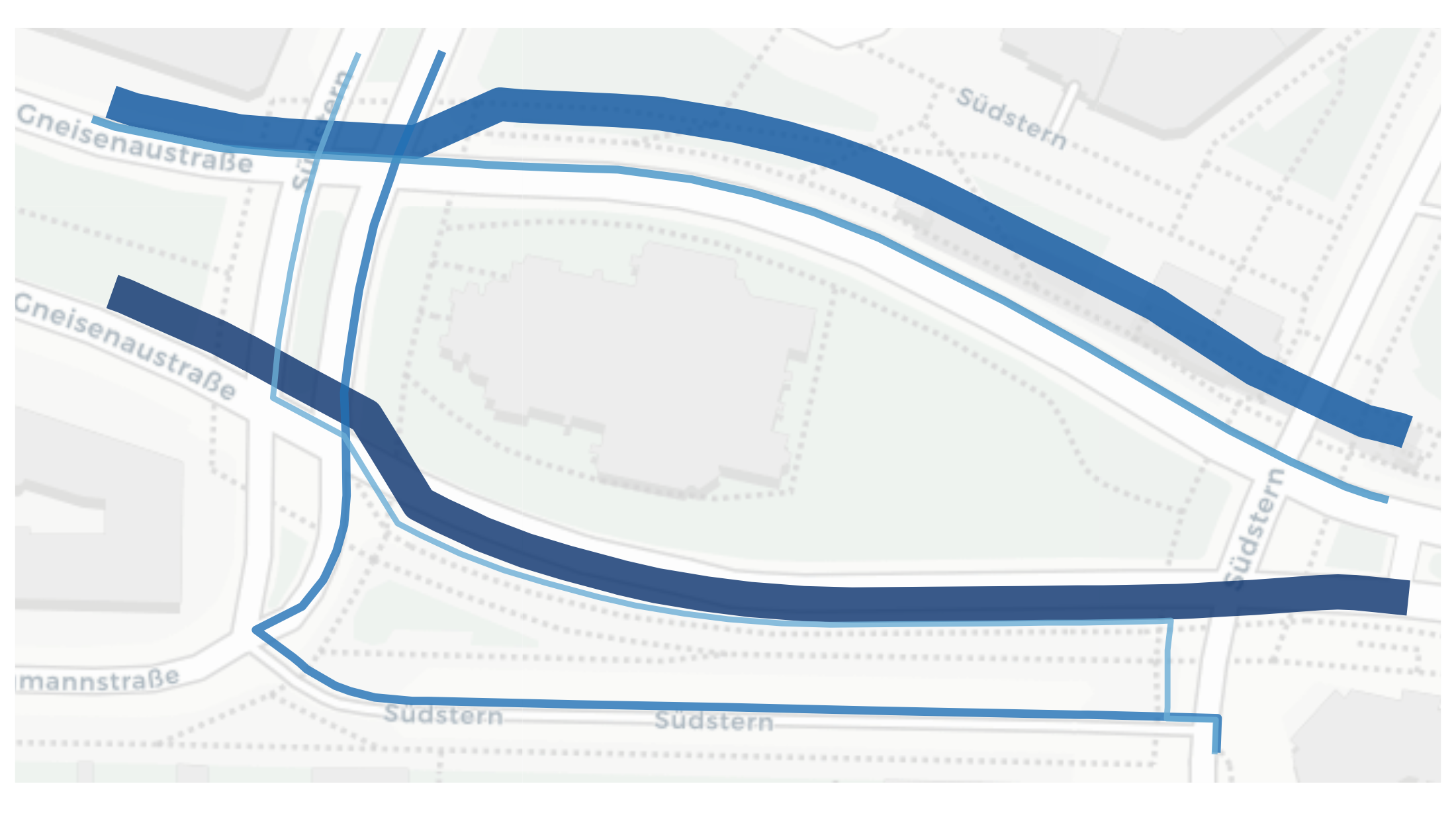}
    \end{minipage}
    \begin{minipage}[t]{0.11\textwidth}
        \includegraphics[width=\textwidth]{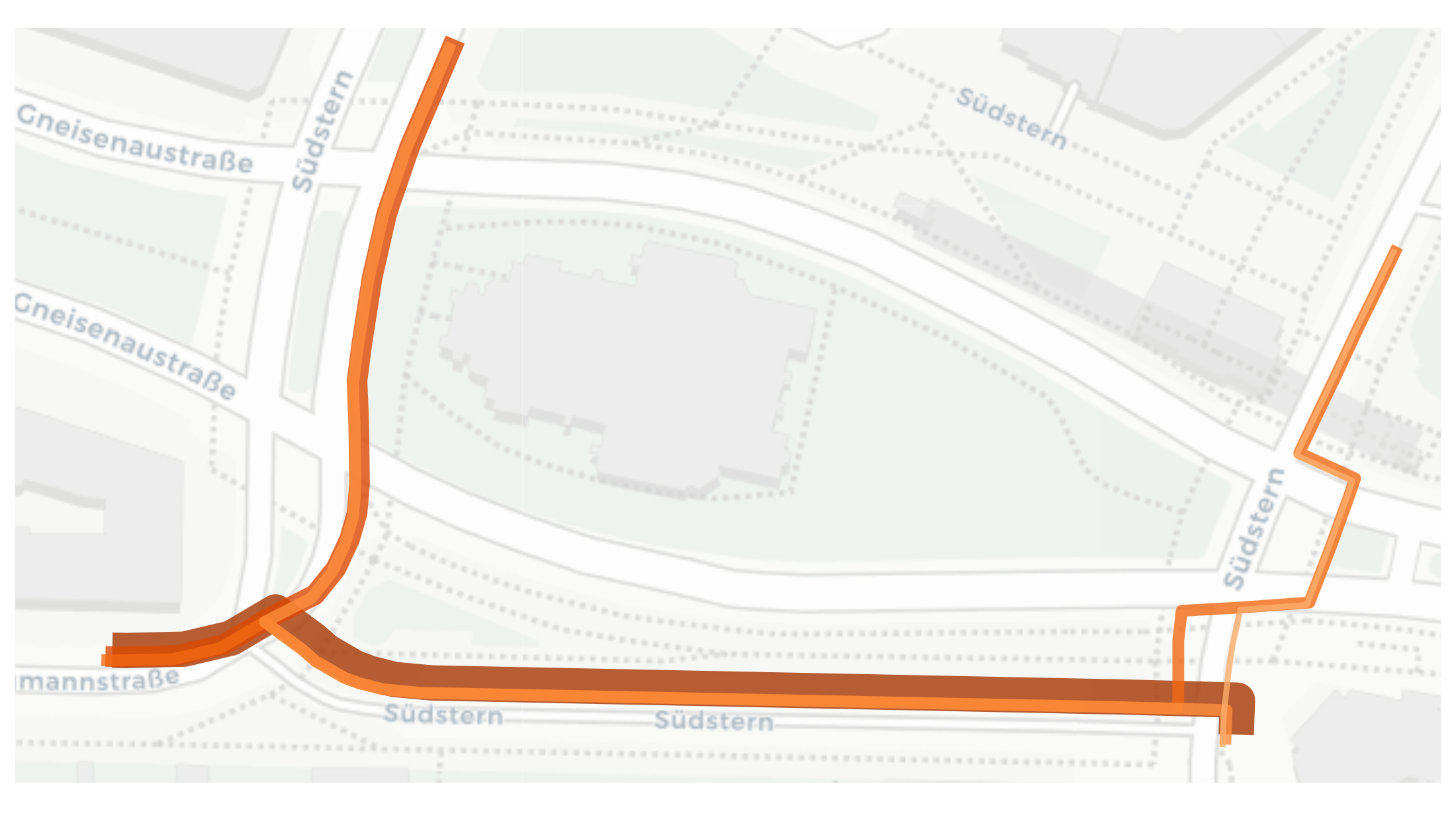}
    \end{minipage}
    \begin{minipage}[t]{0.11\textwidth}
        \includegraphics[width=\textwidth]{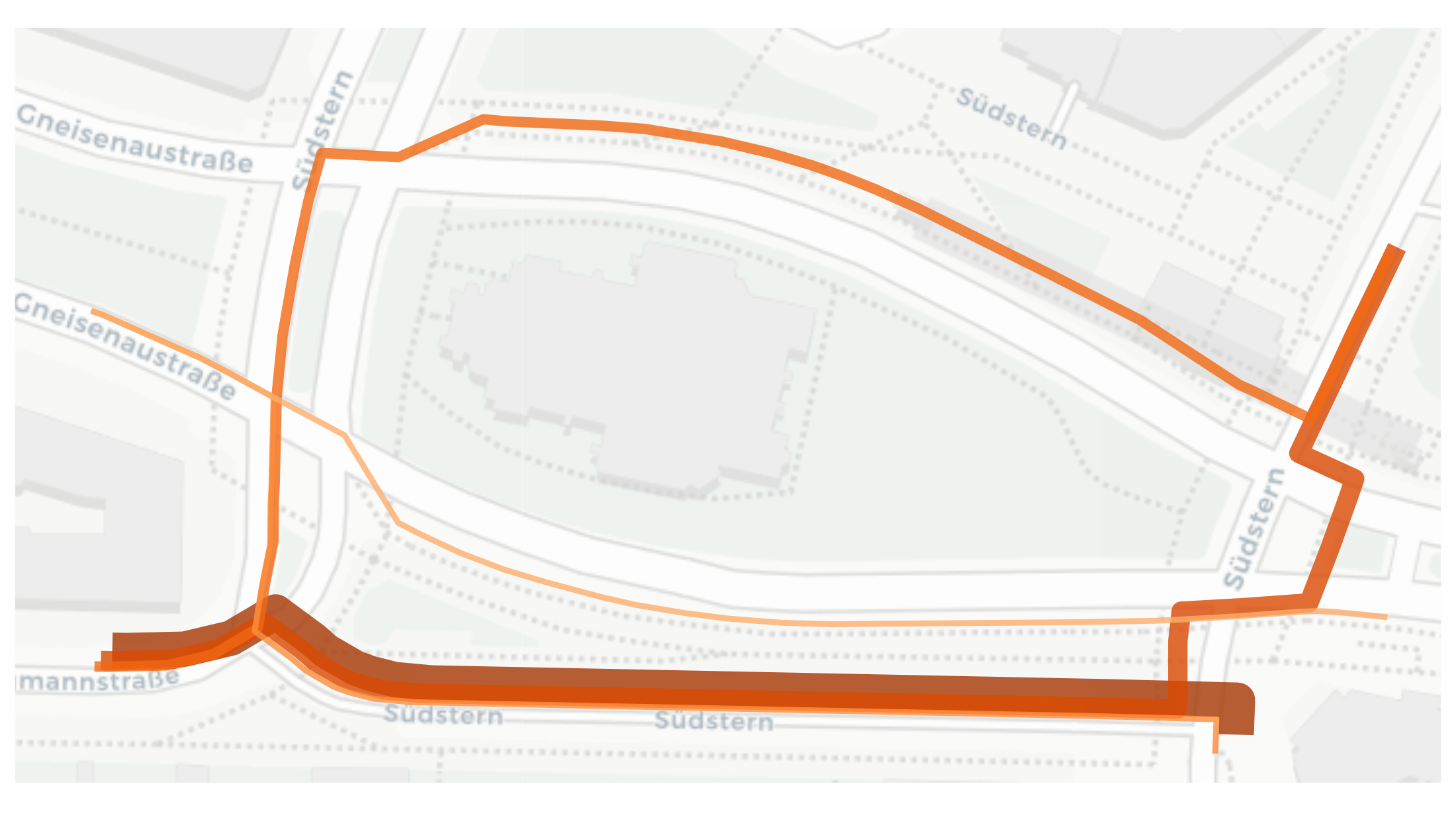}
    \end{minipage}
    \begin{minipage}[t]{0.11\textwidth}
        \includegraphics[width=\textwidth]{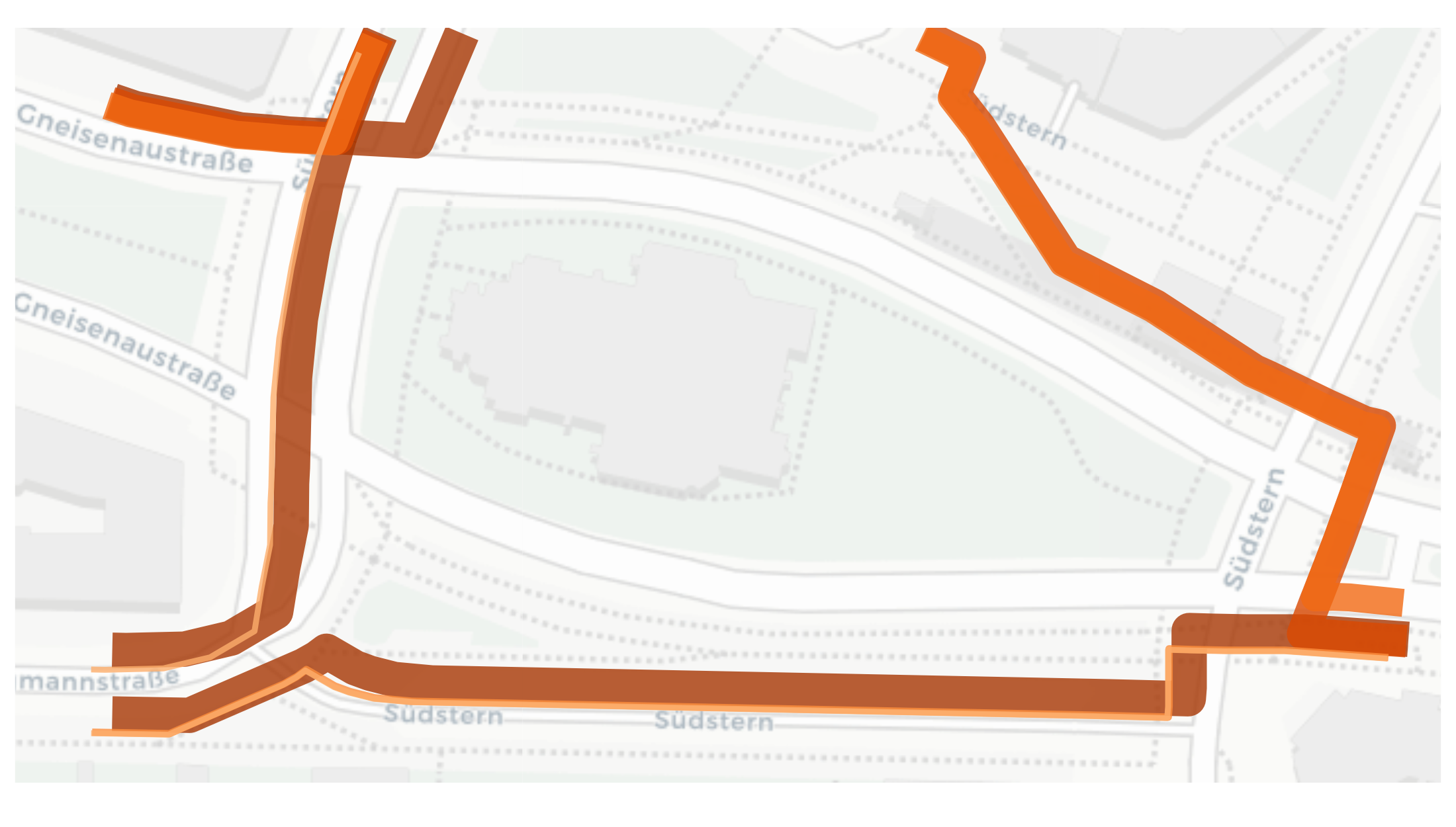}
    \end{minipage}
    \begin{minipage}[t]{0.015\textwidth}
        \rotatebox[origin=l]{90}{\small routed}
    \end{minipage}
    \begin{minipage}[t]{0.11\textwidth}
        \includegraphics[width=\textwidth]{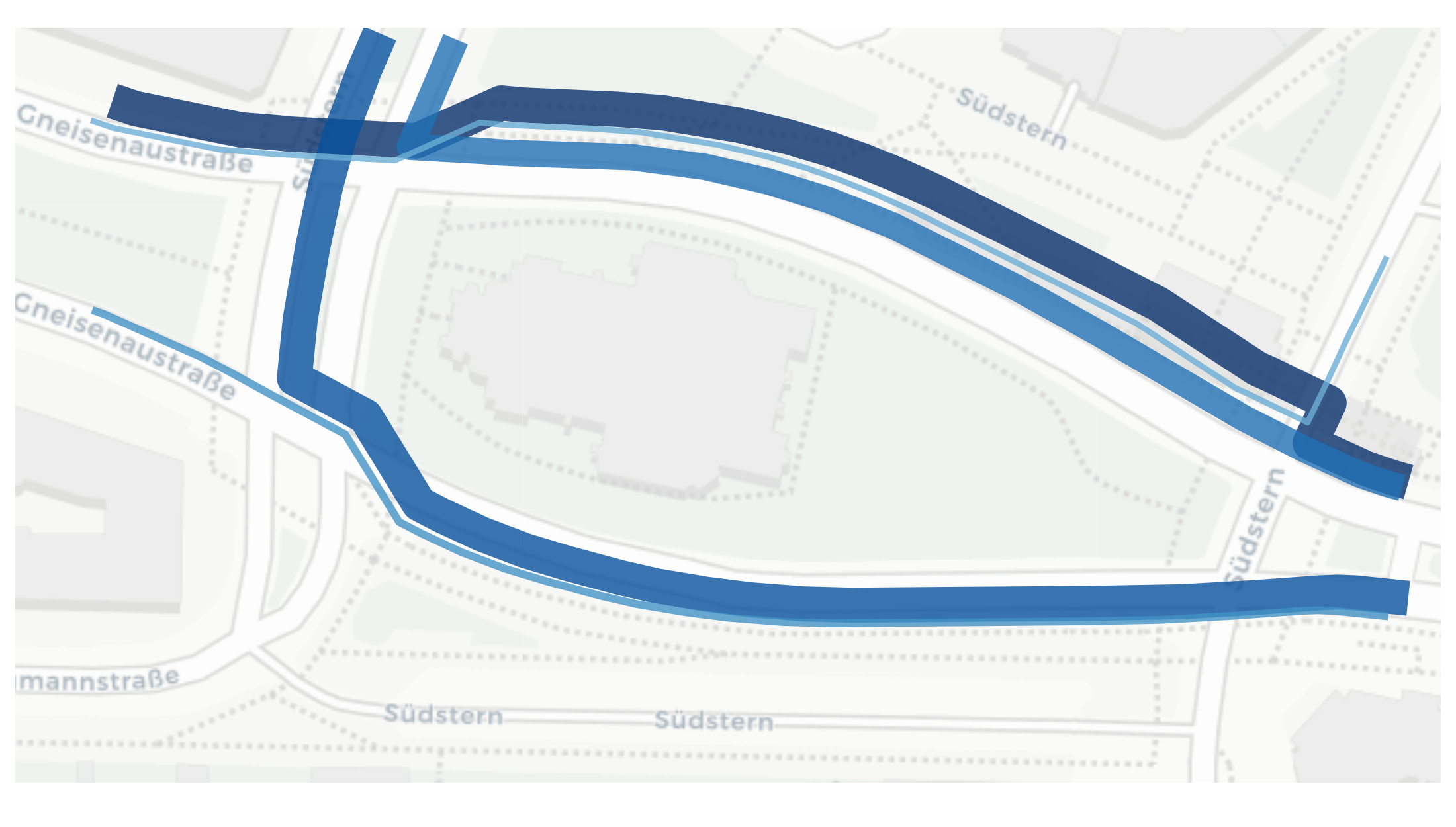}
        \subcaption{raw}
    \end{minipage}
    \begin{minipage}[t]{0.11\textwidth}
        \includegraphics[width=\textwidth]{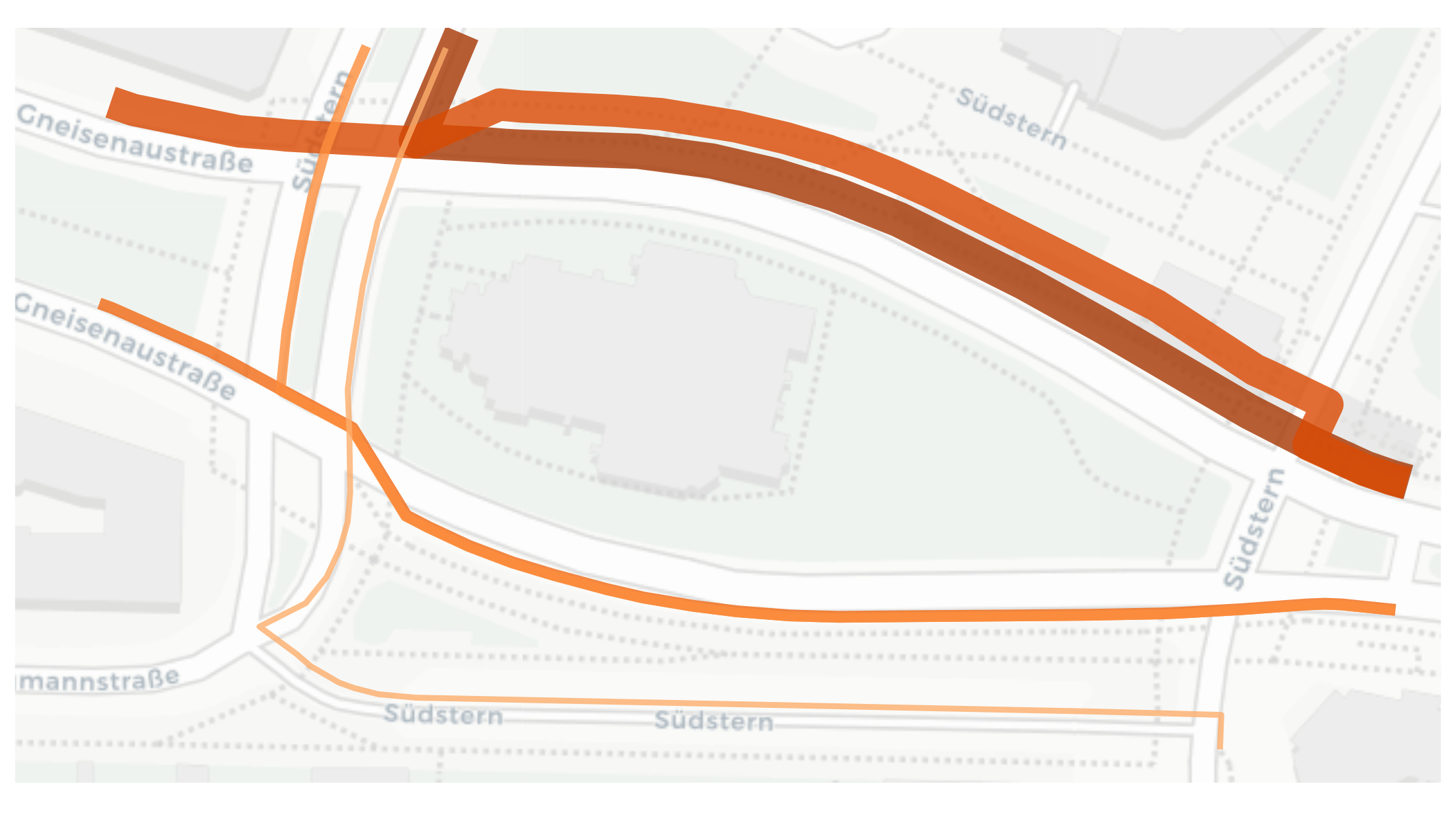}
        \subcaption{AdaTrace}
    \end{minipage}
        \begin{minipage}[t]{0.11\textwidth}
        \includegraphics[width=\textwidth]{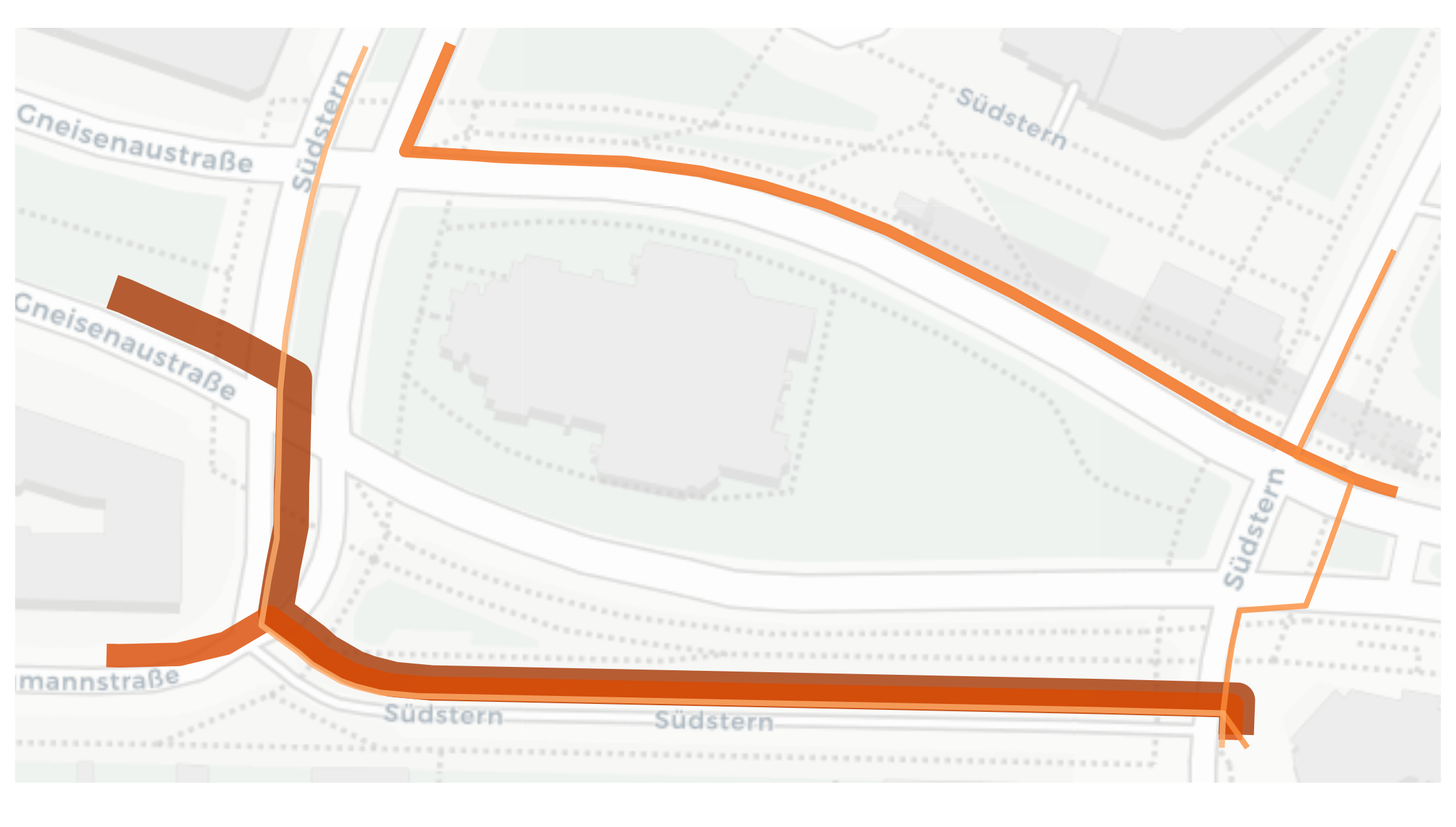}
        \subcaption{PrivTrace}
    \end{minipage}
    \begin{minipage}[t]{0.11\textwidth}
        \includegraphics[width=\textwidth]{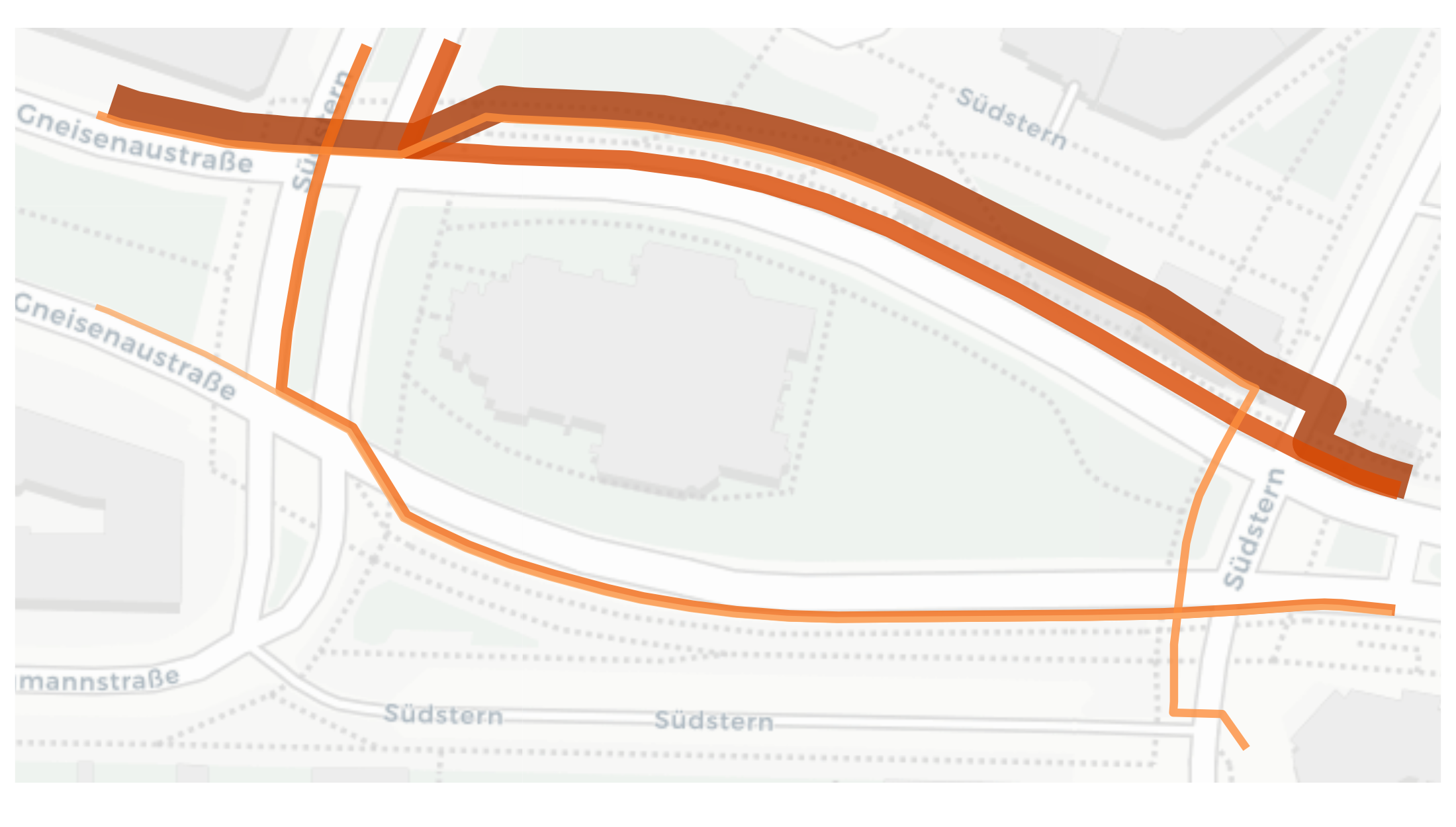}
        \subcaption{BiLSTM}
    \end{minipage}
    \begin{tikzpicture}
        \node[font=\footnotesize] {Basemap: OpenStreetMap contributors};
    \end{tikzpicture}
    \caption{Example intersection showing the top five traffic movements for each matched and routed dataset without DP. Line thickness represents the number of trips.}
    \label{fig:intersectionExample}
\end{figure}

Figure \ref{fig:intersectionExample} shows an example of an intersection displaying the top five traffic movements for each dataset of their matched and routed version.
In general, for the synthetic data to offer added value, the nDCG of the matched version should surpass that of the routed version.  
In Table \ref{tab:intersections}, the averaged results across all intersections are displayed. It can be observed that AdaTrace performs better than its routing baseline for both variations of $k$. Additionally, AdaTrace DP and PrivTrace also exhibit to be slightly better for $k=3$. 

Once again, it is evident that the adaptive grid performs better than the static grid used by BiLSTM. The spatial distribution of matched trips (refer to Appendix \ref{appendix:spatialAgg}) also reflects the visual characteristics of the grid. Although a similar distribution can be observed as in the raw data, there are numerous turns and edges. Employing a finer grid to project the training data could enhance the results.

\begin{table}[tb]
\caption{Averaged nDCG for traffic flow retrieval at 10 intersections, computed for $k=3$ and $k=5$. An nDCG of 1 signifies perfect retrieval. Bold font represents the superior value between the matched and routed versions for each comparison.}
\begin{tabular}{lcc|cc}
\toprule
{} & \multicolumn{2}{c}{k=3} & \multicolumn{2}{c}{k=10} \\
 &  matched & routed &   matched & routed \\
\midrule
raw           &     1.00 &   0.51 &      1.00 &   0.54 \\
\midrule
AdaTrace      &     \textbf{0.58} &   0.47 &     \textbf{ 0.57} &   0.51 \\
AdaTrace(DP)  &    \textbf{0.51} &   0.49 &     \textbf{ 0.53} &   \textbf{0.53} \\
PrivTrace     &     \textbf{0.47} &   0.46 &      0.48 &   \textbf{0.50} \\
PrivTrace(DP) &     0.34 &   \textbf{0.37} &      0.36 &   \textbf{0.37} \\
BiLSTM        &     0.34 &   \textbf{0.49} &      0.39 &   \textbf{0.51} \\
\bottomrule
\end{tabular}
\label{tab:intersections}
\end{table}

\begin{figure}[tb]
    \centering
    \begin{minipage}[t]{0.156\textwidth}
        \includegraphics[width=\textwidth]{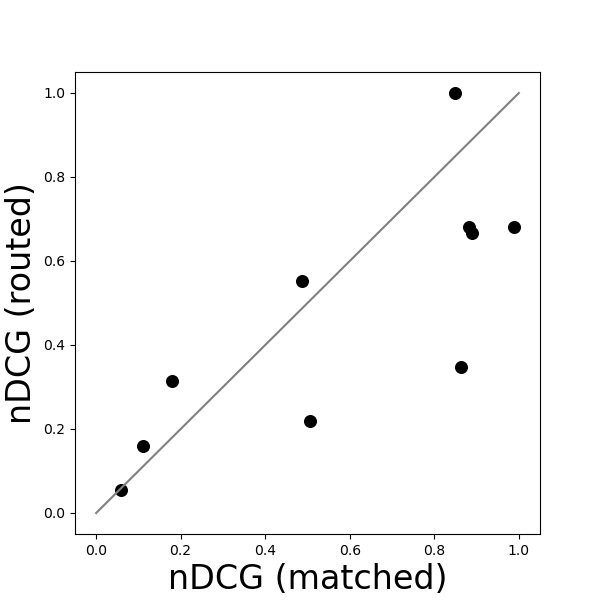}
        \subcaption{AdaTrace}
    \end{minipage}  
    \begin{minipage}[t]{0.156\textwidth}
        \includegraphics[width=\textwidth]{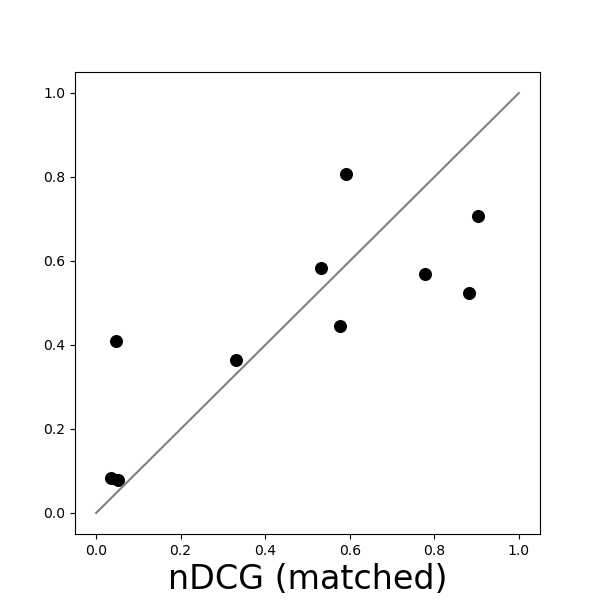}
        \subcaption{PrivTrace}
    \end{minipage}  
    \begin{minipage}[t]{0.156\textwidth}
        \includegraphics[width=\textwidth]{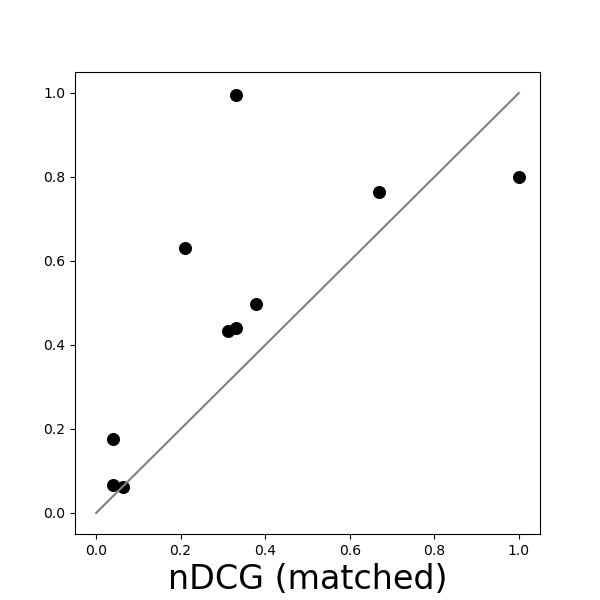}
        \subcaption{BiLSTM}
    \end{minipage}  
    \caption{Traffic flow retrieval according to nDCG with $k=3$: each point corresponds to one intersection for selected models. Points above (below) the diagonal indicate a superior routed (matched) value.}
    \label{fig:intersecScatter}
\end{figure}

To assess the practical significance of the results, we analyzed the distribution of nDCG values, visualized in Figure \ref{fig:intersecScatter}. 
The routed version of BiLSTM outperforms the matching on almost all intersections. 
Even for the top-performing AdaTrace, matching outperforms routing only in 6 out of 10 instances. Additionally, in half of the cases, the nDCG scores for `matched' results are equal to or below 0.5. These findings highlight that while certain intersections may exhibit accurate traffic flow representation, there remains a 50\% likelihood of erroneous outcomes. Consequently, the practical utility of the results becomes limited.

\section{Discussion and Conclusion}
\label{sec:discussion}

The potential of synthetic data lies in its (promised) ability to provide full flexibility in analyses and modeling endeavors while ensuring a high level of privacy. In our research, we delved into the practical implications of `full flexibility’ for trip data consisting of fine-granular routes as typically recorded using GPS devices. 

To address the question of what constitutes high utility for trip data and how it can be measured (RQ1), we assert that the metrics employed should prioritize the detailed, sequential nature of trip data. Our proposal suggests evaluating the  lengths of trips, traffic volumes, preferences for specific road segments, and the flow of traffic at intersections (in addition to standard high-level metrics already used in the literature). These measurements (along with the utilization of synthetic trip data at the street level) necessitate an additional computationally intensive step known as `map matching'. We further contend that synthetic data must surpass those generated by standard routing algorithms in order to claim a significant advantage for subsequent analyses. 

It is worth noting that the models examined in this study, similar to the majority of models used for generating synthetic mobility data, largely exclude attributes that are crucial for real-life applications, such as timestamps, traffic mode, and user-level information. This particularly includes the inability to group trips per user. This limitation restricts the potential applications of synthetic data to spatial information alone, which prompted the consideration of the aforementioned metrics.

We evaluated the utility of five state-of-the-art models, AdaTrace, PrivTrace, DP-Loc, a BiLSTM-based model, and TrajGAIL, using the designated utility metrics on a dataset comprising  approximately 30,000 bicycle trips in Berlin. 
Regarding RQ2, we observed that TrajGAIL failed to generate data within a reasonable computation time for city-scale scenarios, while DP-Loc frequently generated jumps that were unsuitable for map matching. The remaining three models  maintained a certain degree of spatial distribution, as evidenced by the analysis of road preferences. AdaTrace showed overall the best results. It allowed survey participants to identify preferred roads with 90\% accuracy and an F1-score of $\geq 0.7$. Additionally, both AdaTrace and PrivTrace slightly outperformed the routing baseline in  analyzing traffic volumes. The adaptive grid employed likely succeeds in maintaining `hotspots' of the spatial distribution.
The BiLSTM-based model might potentially yield superior 
results if trained on a grid with a finer resolution and a sufficiently large dataset. However, the resulting increase in computational cost raises concerns regarding its practical feasibility. Overall, the BiLSTM results remain  inconclusive since, contrary to the findings regarding traffic volumes on road segments, it performed best in terms of statistical similarity of road preferences and achieved  results similar to AdaTrace in human classification of road preferences. This underscores the need to consider multiple metrics 
as a reliable measure of the success of arbitrary downstream tasks related to the respective distribution.

Additionally, we found unsatisfactory utility when evaluating the ratio of trip length to straight-line distance, with clear superiority of the routing baseline for all models. With respect to traffic flow at intersections, only AdaTrace managed to surpass the routing baseline substantially among the evaluated models, but even then, the provided level of utility remained questionable.

With regards to (RQ3), only AdaTrace and PrivTrace generated useful differentially private data. 
As the application of a BiLSTM without DP raises major privacy issues \cite{luca_survey_2021}, further improvements are needed for it to serve as a sensible privacy technique.
Notably, despite increasing AdaTrace's spatial resolution from the default setting of $6x6$ to a more-detailed $28x28$ top-level grid, the DP version still provided superior results compared to PrivTrace without DP. It also outperformed the BiLSTM model in two analyses and was only slightly inferior in two other evaluations. However, the practical utility remains doubtful.
In addition, it should be noted that all evaluated models only offer item-level DP as user IDs are discarded. This raises concerns about the provided privacy level, particularly if only few power users contribute a significant portion of the data. 

We further acknowledge that there may be more suitable map matching algorithms for synthetic data than the utilized OSRM implementation. However, we rather suggest the advancement of synthesis algorithms such that road networks are inherently considered instead of optimizing post-processing techniques.

The SimRa dataset employed in this study is of significant importance from an urban planning point of view, given its representation of a substantial number of bicycle trips and  riders' perceived safety. 
While the evaluation on other datasets that differ regarding size, city structure or traffic modes might vary to a certain extent, we expect the overarching trends to remain consistent.
We aim to broaden our analyses in subsequent research to further explore the models' suitability, especially for taxi trips and larger datasets.

In summary, our results raise the question of whether models claiming high flexibility truly provide sufficient benefits or, on the contrary, may do more harm by failing to deliver reliable results. 
If such models merely manage to maintain a moderate level of spatial granularity, there might be more accurate and privacy-preserving methods available through the provision of aggregate data.
It is likely that generating synthetic data offering both high flexibility and strong privacy guarantees is not possible \cite{Stadler:298904}. Instead of striving for full flexibility, it is advisable to clearly indicate the specific applications for which the respective model is well-suited and those for which it is not.
For instance, synthetic trip data could prove valuable  for development purposes or to gain a preliminary understanding of raw data before utilizing it in a controlled and secure environment for actual analyses. In such cases, the focus may be on maintaining the sampling rate and accuracy of GPS data rather than precisely mimicking the actual spatial distributions. For such applications, it may be more important to retain all attributes instead of solely focusing on spatial information.

\bibliographystyle{ACM-Reference-Format}
\bibliography{references}

\pagebreak

\section*{Appendix}
\subsection{Map matching validity}
\label{appendix:matching}

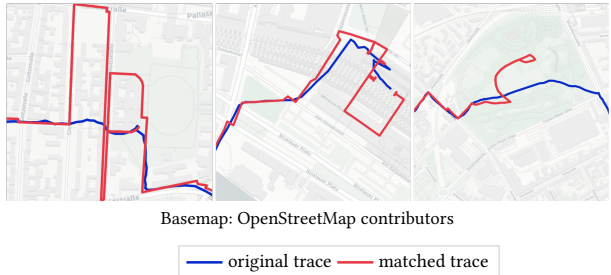
\begin{figure}[h]
    \centering
     \includegraphics[width=8cm]{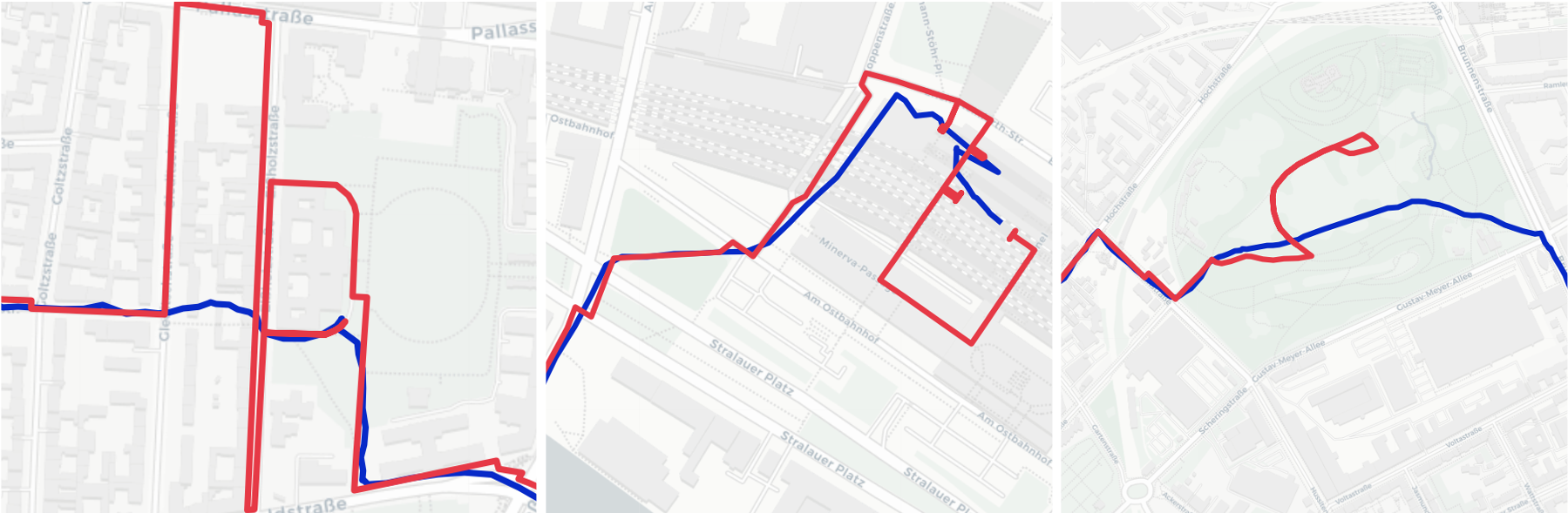}
      \begin{tikzpicture}
            \node[font=\footnotesize] {Basemap: OpenStreetMap contributors};
      \end{tikzpicture}
       \input{figures/matching_fails_legend}
        \vspace*{-4cm}
    \caption{Examples of map matching inaccuracies. \textit{Left}: A cyclist used a footpath that was matched to a legally accessible path. \textit{Center:} Start or end of the trace is off the road network - in this example, at a train station. \textit{Right:} Matching terminates when there is no nearby path available - here, a cyclist rides on an uncharted path in a park.}

    \label{fig:mismatches}
\end{figure}

There are cases in which the matching API is not suitable for capturing actually traveled bicycle routes. Figure \ref{fig:mismatches} provides examples of potential issues.

Also, it is possible for the API to entirely  fail in producing a matched trace. A solid map matching is crucial for ensuring the validity of our results. Thus, 
we assess the frequency of complete  matching failures  and matched trips shorter than 90\% of their original counterpart, indicating  termination in the midst of the trace.
Table \ref{tab:failedMatches} suggests that for all datasets, matching entirely fails only for at most $4\%$ of trips and partly for $18\%$, which is acceptable for the purposes of our evaluation.

\begin{table}[h]
    \caption{Percentage of completely and partly failed  matches per dataset.}
    \label{tab:failedMatches}
    \centering
\begin{tabular}{lrr}
\toprule
    &  failed (\%) &  shorter (\%) \\
\midrule
raw &         0 &          7 \\
\midrule
AdaTrace &           1 &           5 \\
AdaTrace DP &           1 &           13 \\
PrivTrace &           2 &           6 \\
PrivTrac DP &           4 &          5 \\
BiLSTM &          0 &          18 \\
\bottomrule
\end{tabular}

\end{table}

To further assess the validity of the chosen matching algorithm, we evaluate the map matching using the Hausdorff distance.
 In an accurate map matching scenario, the Hausdorff distance should be small, indicating that no points in the matched trip are far away from their corresponding point in the original trip.
 As expected for successful map matching, all synthetic datasets show a Hausdorff distance in the order of magnitude of the selected grid resolution, and the raw data has a median Hausdorff distance of only $33m$.

\subsection{Original dataset visualization}
\label{appendix:vizOriginal}

\begin{figure}[]
    \centering
     \begin{minipage}[t]{0.49\textwidth}
        \includegraphics[width=0.49\textwidth]{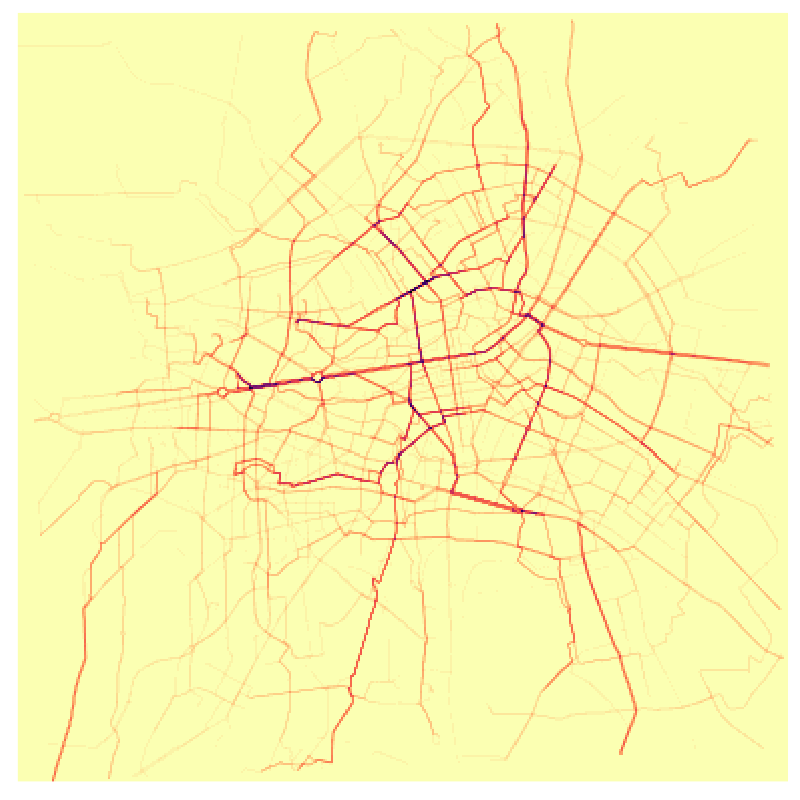}
        \includegraphics[width=0.49\textwidth]{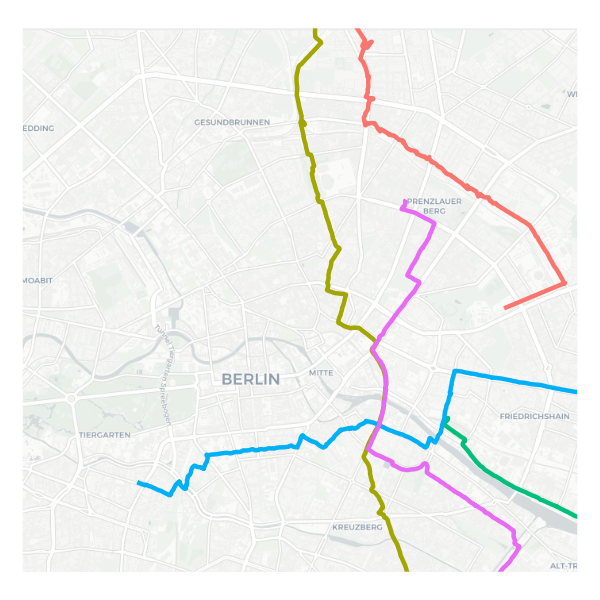}
        \subcaption{raw}
    \end{minipage}
    \begin{minipage}[t]{0.49\textwidth}
        \includegraphics[width=0.49\textwidth]{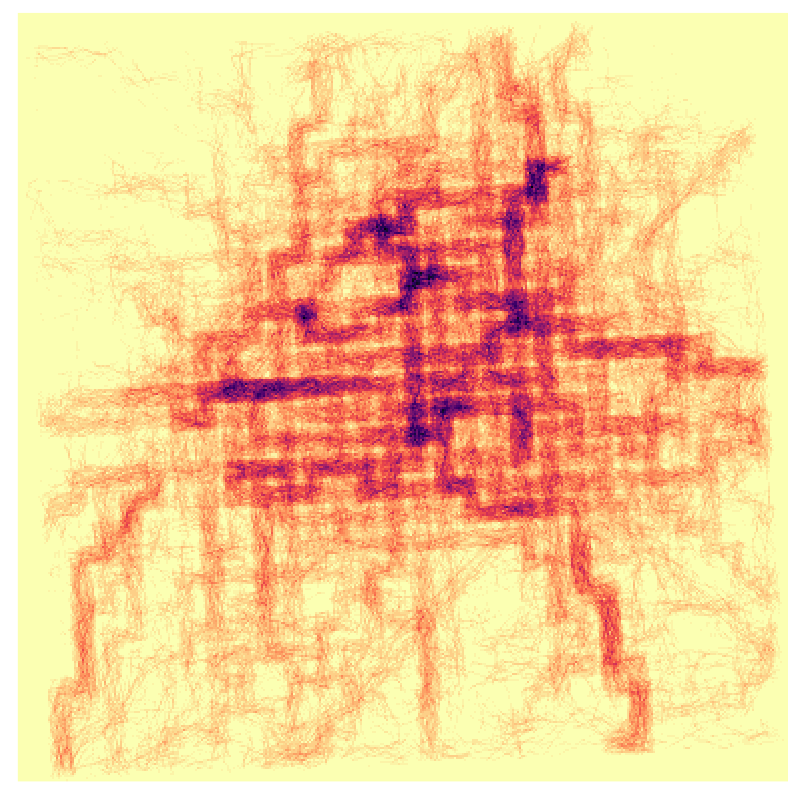}
        \includegraphics[width=0.49\textwidth]{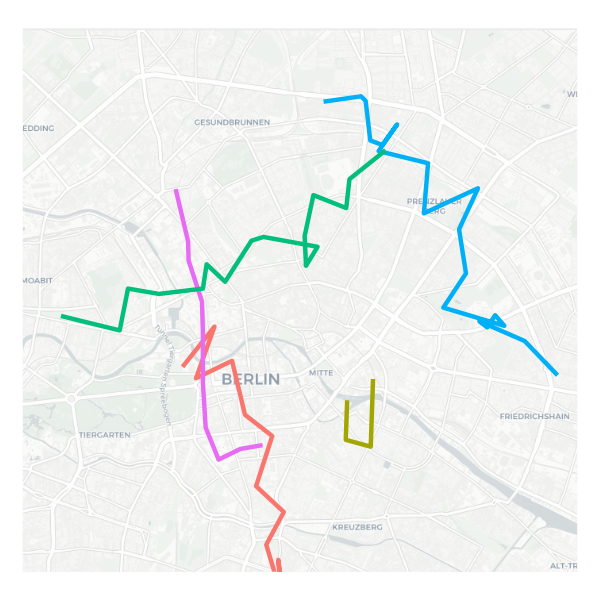}
        \subcaption{AdaTrace}
    \end{minipage}
    \begin{minipage}[t]{0.49\textwidth}
        \includegraphics[width=0.49\textwidth]{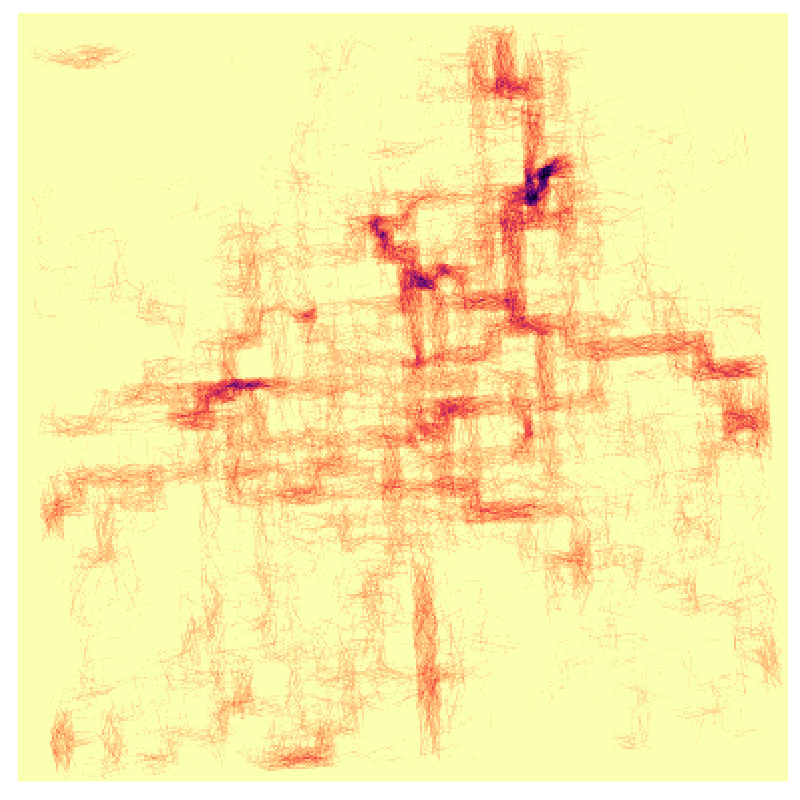}
        \includegraphics[width=0.49\textwidth]{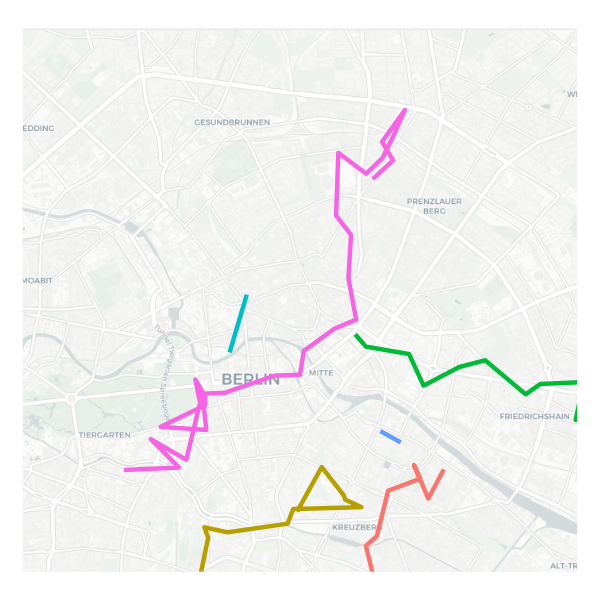}
        \subcaption{PrivTrace}
    \end{minipage}
    \begin{minipage}[t]{0.49\textwidth}
        \includegraphics[width=0.49\textwidth]{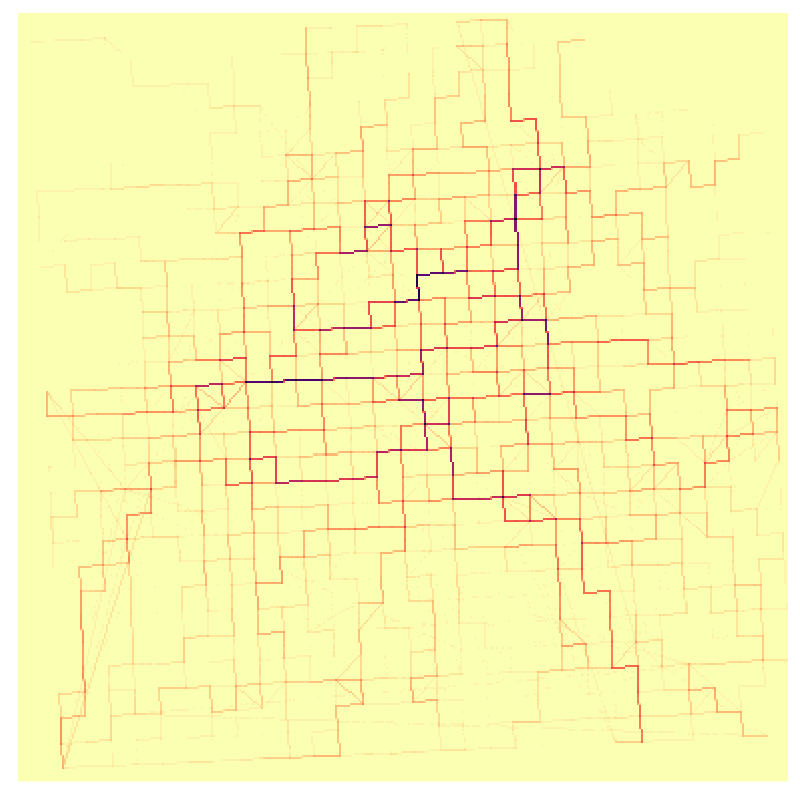}
        \includegraphics[width=0.49\textwidth]{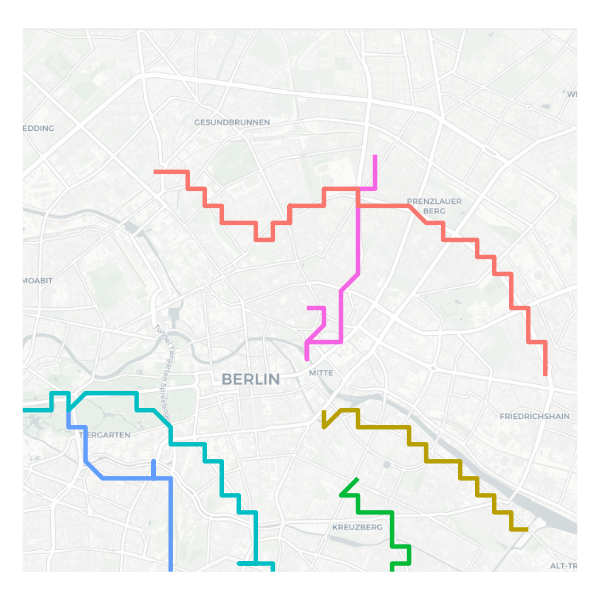}
        \subcaption{BiLSTM}
    \end{minipage}
    \begin{tikzpicture}
    \node[font=\footnotesize] {Basemap: OpenStreetMap contributors};
  \end{tikzpicture}
    \caption{Spatial distribution (left) of original datasets without DP and respective example trips (right).}
    \label{fig:originalGrid}
\end{figure}
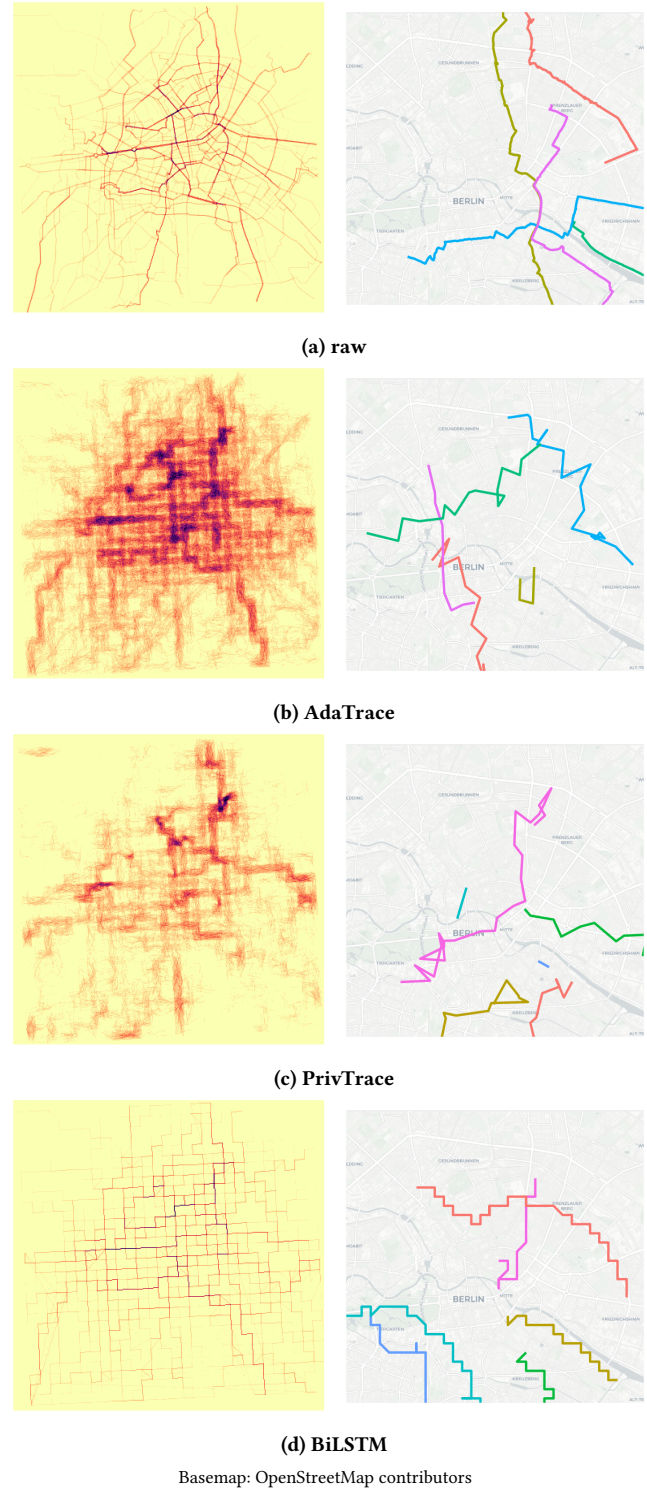

Figure \ref{fig:originalGrid} visualizes the original datasets without DP with respect to their spatial distribution 
as well as example trips.

\subsection{Matched and routed spatial distributions}
\label{appendix:spatialAgg}

\begin{figure}[b]
    \centering
     \begin{minipage}[t]{0.48\textwidth}
     \centering
        \includegraphics[width=0.48\textwidth]{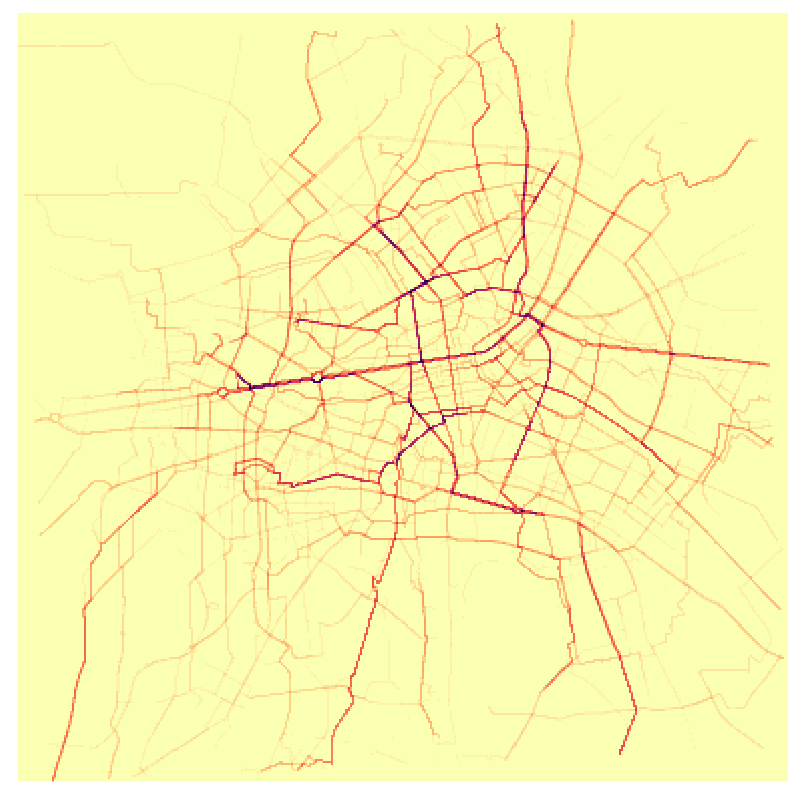}
        \includegraphics[width=0.48\textwidth]{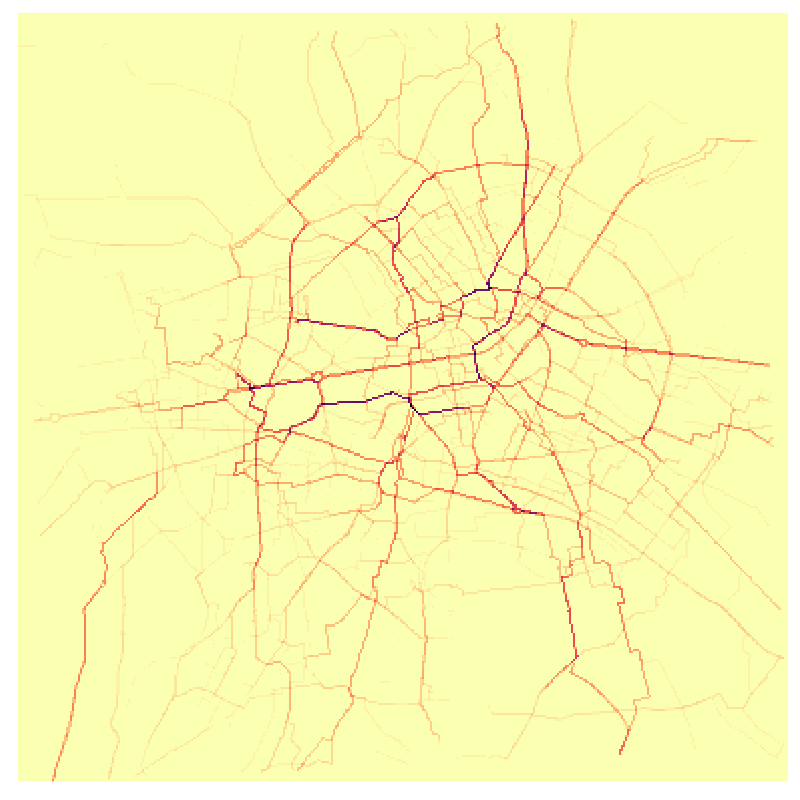}
        \subcaption{raw}
    \end{minipage}
     \begin{minipage}[t]{0.48\textwidth}
         \centering
        \includegraphics[width=0.48\textwidth]{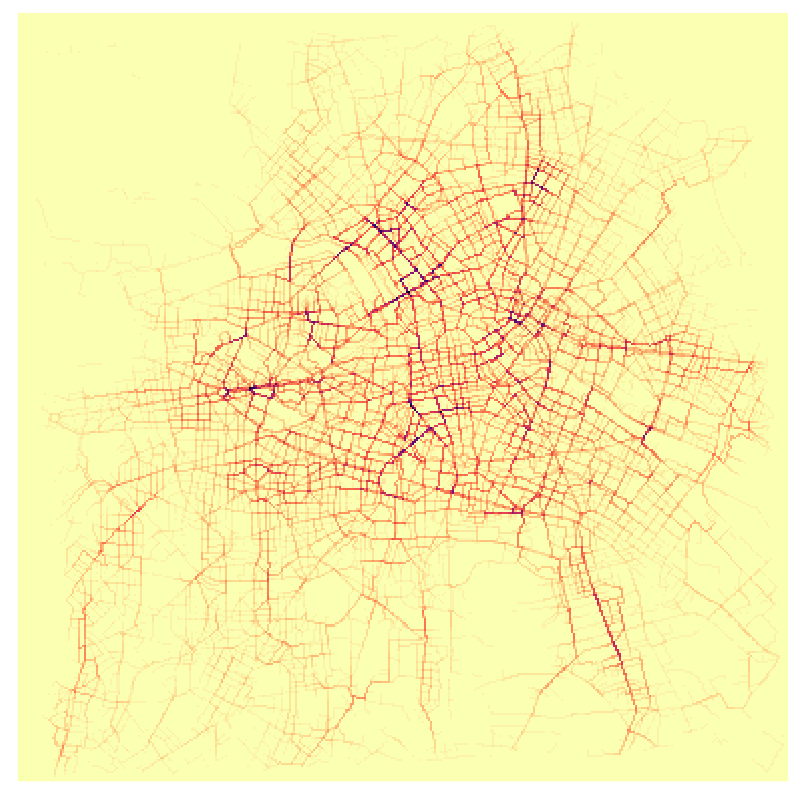}
        \includegraphics[width=0.48\textwidth]{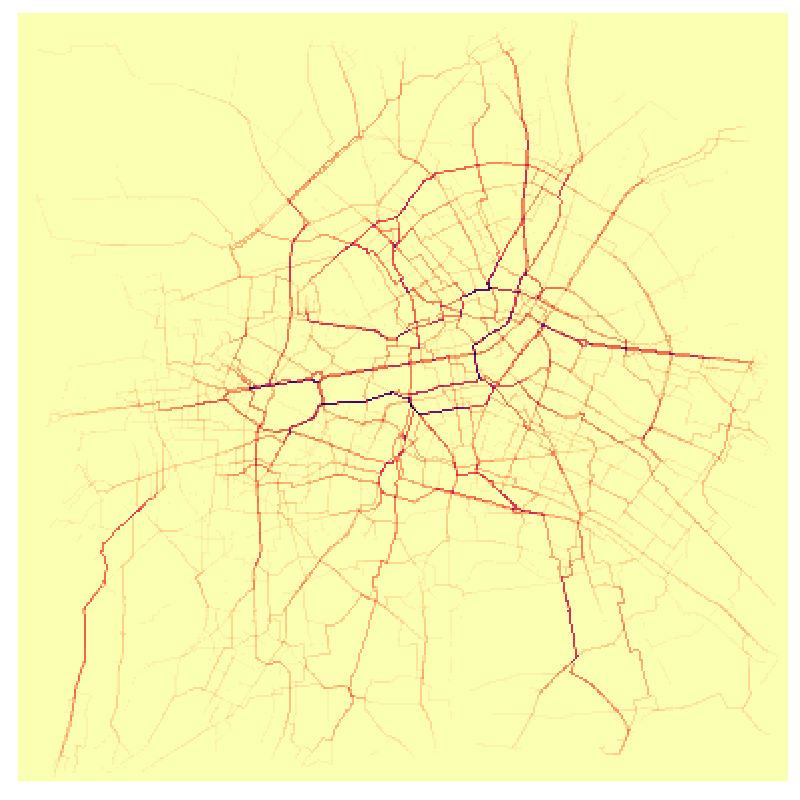}
        \subcaption{AdaTrace}
    \end{minipage}
     \begin{minipage}[t]{0.48\textwidth}
             \centering
        \includegraphics[width=0.48\textwidth]{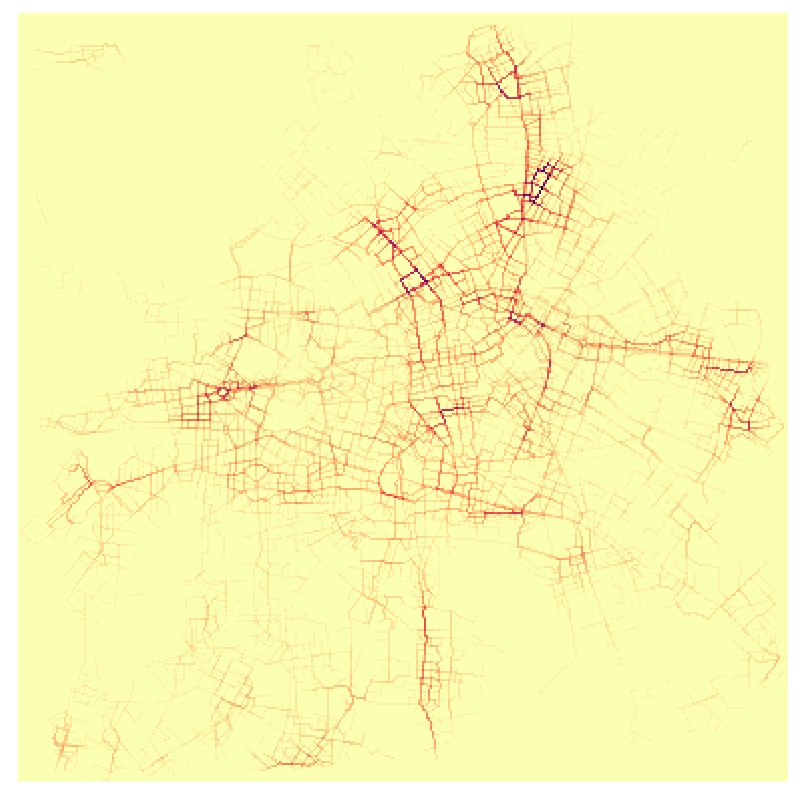}
        \includegraphics[width=0.48\textwidth]{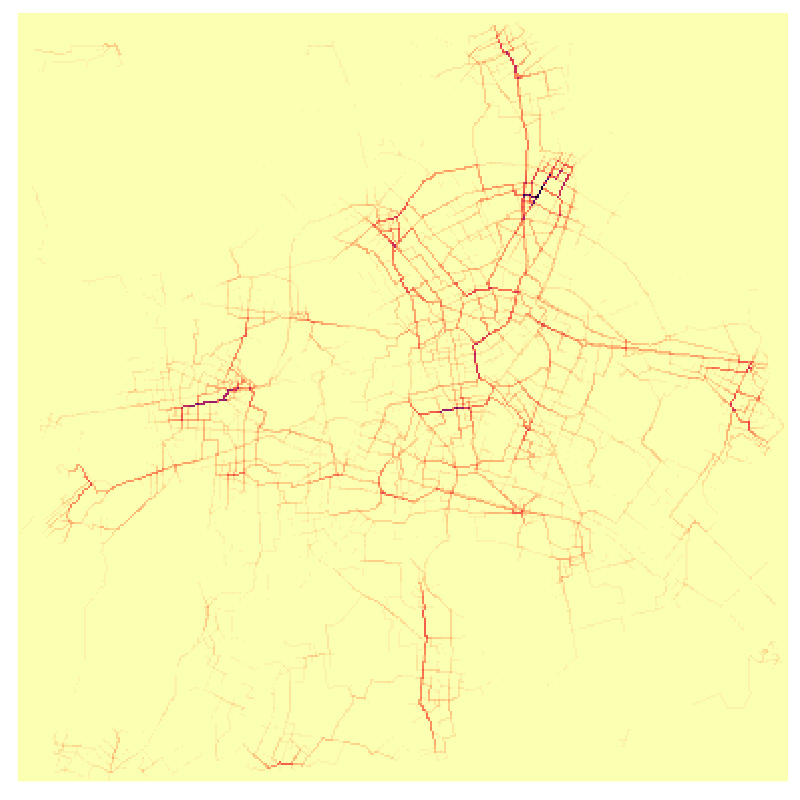}
        \captionsetup{skip=1pt} 
        \subcaption{PrivTrace}
    \end{minipage}
     \begin{minipage}[t]{0.48\textwidth}
         \centering
        \includegraphics[width=0.48\textwidth]{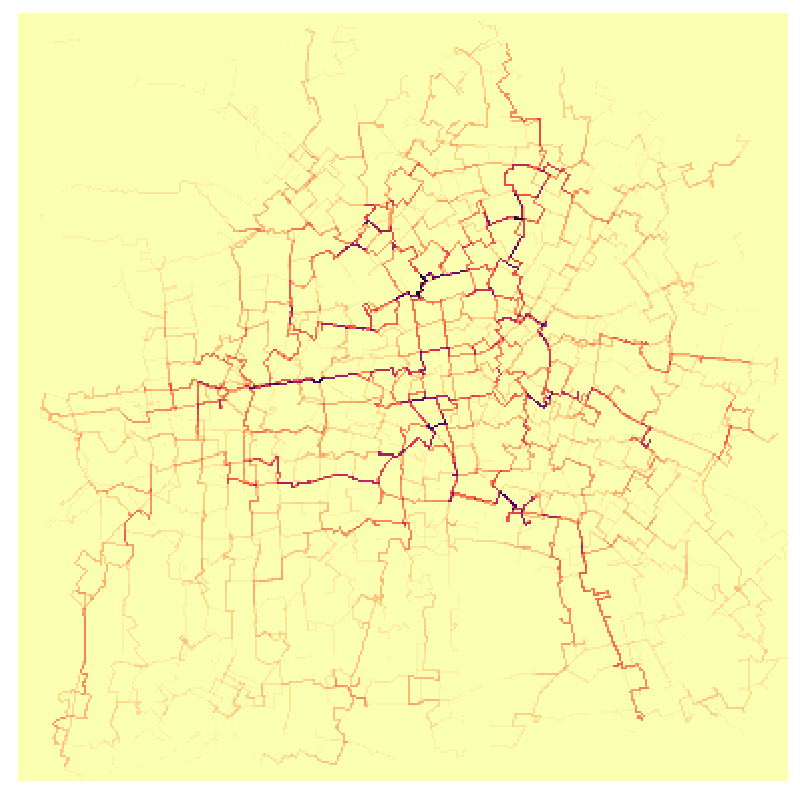}
        \includegraphics[width=0.48\textwidth]{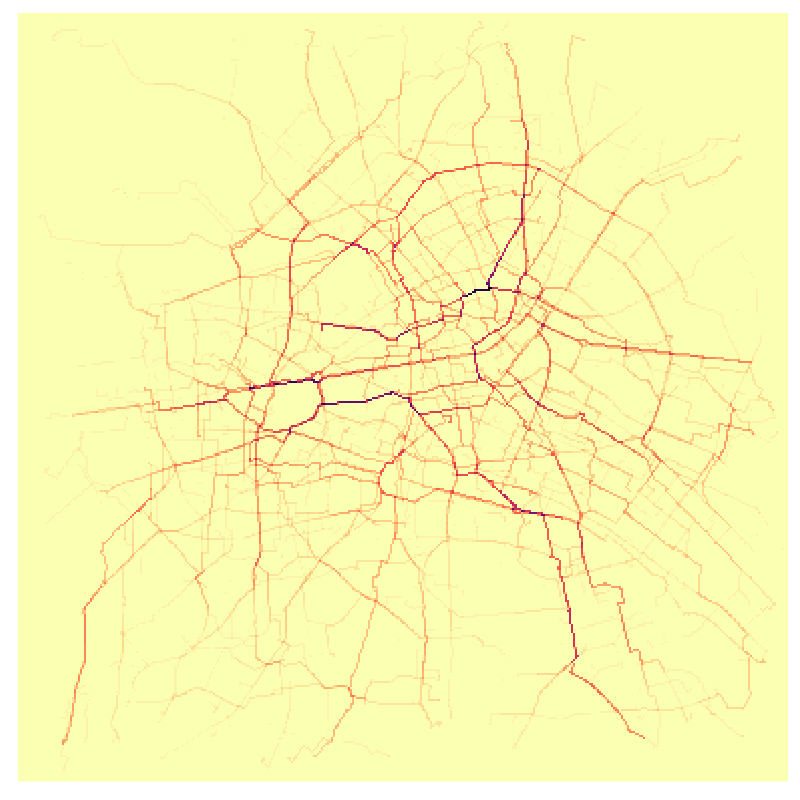}
        \subcaption{BiLSTM}
    \end{minipage}
        \caption{Spatial distribution of matched (left) and routed (right) datasets without DP.}
    \label{fig:spatAgg}
\end{figure}

Figure \ref{fig:spatAgg} shows spatial distributions for each dataset without DP in matched and routed versions. 

\pagebreak

\end{document}

%% file: figures/matching_legend.tex
\begin{tikzpicture}[scale=0.8]

	\definecolor{color0}{RGB}{7,42,200}
	\definecolor{color1}{RGB}{230,57,70}
	\definecolor{color2}{RGB}{255,189,0}
	\definecolor{color3}{RGB}{108,117,125}

    \begin{axis}[
		hide axis,
		xmin=0,
		xmax=1,
		ymin=0,
		ymax=1,		
		legend cell align={left},
		legend columns=4,
		legend style={
			draw opacity=1,
			text opacity=1,
			draw=white!80!black
		}
		]
		\addlegendentry{original}
		\addlegendimage{very thick, color0}
		\addlegendentry{matched}
		\addlegendimage{very thick, color1}
  		\addlegendentry{routed}
		\addlegendimage{very thick, color2}
		\addlegendentry{straight-line}
		\addlegendimage{very thick, color3}
\end{axis}

\end{tikzpicture}

%% file: figures/matching_fails_legend.tex
\begin{tikzpicture}[scale=0.8]

	\definecolor{color0}{RGB}{7,42,200}
	\definecolor{color1}{RGB}{230,57,70}

    \begin{axis}[
		hide axis,
		xmin=0,
		xmax=1,
		ymin=0,
		ymax=1,		
		legend cell align={left},
		legend columns=3,
		legend style={
			draw opacity=1,
			text opacity=1,
			draw=white!80!black
		}
		]
		\addlegendentry{original trace}
		\addlegendimage{very thick, color0}
		\addlegendentry{matched trace}
		\addlegendimage{very thick, color1}
        \end{axis}

\end{tikzpicture}

%% file: main.bbl

\begin{thebibliography}{23}


\ifx \showCODEN    \undefined \def \showCODEN     #1{\unskip}     \fi
\ifx \showDOI      \undefined \def \showDOI       #1{#1}\fi
\ifx \showISBNx    \undefined \def \showISBNx     #1{\unskip}     \fi
\ifx \showISBNxiii \undefined \def \showISBNxiii  #1{\unskip}     \fi
\ifx \showISSN     \undefined \def \showISSN      #1{\unskip}     \fi
\ifx \showLCCN     \undefined \def \showLCCN      #1{\unskip}     \fi
\ifx \shownote     \undefined \def \shownote      #1{#1}          \fi
\ifx \showarticletitle \undefined \def \showarticletitle #1{#1}   \fi
\ifx \showURL      \undefined \def \showURL       {\relax}        \fi
\providecommand\bibfield[2]{#2}
\providecommand\bibinfo[2]{#2}
\providecommand\natexlab[1]{#1}
\providecommand\showeprint[2][]{arXiv:#2}

\bibitem[Abadi et~al\mbox{.}(2016)]%
        {abadi_deep_2016}
\bibfield{author}{\bibinfo{person}{Martin Abadi}, \bibinfo{person}{Andy Chu},
  \bibinfo{person}{Ian Goodfellow}, \bibinfo{person}{H.~Brendan McMahan},
  \bibinfo{person}{Ilya Mironov}, \bibinfo{person}{Kunal Talwar}, {and}
  \bibinfo{person}{Li Zhang}.} \bibinfo{year}{2016}\natexlab{}.
\newblock \showarticletitle{Deep {{Learning}} with {{Differential Privacy}}}.
  In \bibinfo{booktitle}{\emph{Proceedings of the 2016 {{ACM SIGSAC
  Conference}} on {{Computer}} and {{Communications Security}}}}
  \emph{(\bibinfo{series}{{{CCS}} '16})}. \bibinfo{publisher}{{Association for
  Computing Machinery}}, \bibinfo{address}{{New York, NY, USA}},
  \bibinfo{pages}{308--318}.
\newblock
\showISBNx{978-1-4503-4139-4}
\urldef\tempurl%
\url{https://doi.org/10.1145/2976749.2978318}
\showDOI{\tempurl}


\bibitem[Bermbach(2021)]%
        {bermbach_simra_2021}
\bibfield{author}{\bibinfo{person}{David Bermbach}.}
  \bibinfo{year}{2021}\natexlab{}.
\newblock \showarticletitle{{{SimRa Rides Berlin}} 06/19 - 12/20}.
\newblock  (\bibinfo{date}{Feb.} \bibinfo{year}{2021}).
\newblock
\urldef\tempurl%
\url{https://doi.org/10.14279/DEPOSITONCE-10605}
\showDOI{\tempurl}


\bibitem[Bermbach and Karakaya(2021)]%
        {bermbach_simra_2021-1}
\bibfield{author}{\bibinfo{person}{David Bermbach} {and}
  \bibinfo{person}{Ahmet-Serdar Karakaya}.} \bibinfo{year}{2021}\natexlab{}.
\newblock \showarticletitle{{{SimRa Rides Berlin}} 01/21 - 09/21}.
\newblock  (\bibinfo{date}{Oct.} \bibinfo{year}{2021}).
\newblock
\urldef\tempurl%
\url{https://doi.org/10.14279/DEPOSITONCE-12452}
\showDOI{\tempurl}


\bibitem[{Blanco-Justicia} et~al\mbox{.}(2022)]%
        {blanco-justicia_generation_2022}
\bibfield{author}{\bibinfo{person}{Alberto {Blanco-Justicia}},
  \bibinfo{person}{Najeeb~Moharram Jebreel}, \bibinfo{person}{Jes{\'u}s~A.
  Manj{\'o}n}, {and} \bibinfo{person}{Josep {Domingo-Ferrer}}.}
  \bibinfo{year}{2022}\natexlab{}.
\newblock \showarticletitle{Generation of~{{Synthetic Trajectory Microdata}}
  from~{{Language Models}}}. In \bibinfo{booktitle}{\emph{Privacy in
  {{Statistical Databases}}}} \emph{(\bibinfo{series}{Lecture {{Notes}} in
  {{Computer Science}}})}, \bibfield{editor}{\bibinfo{person}{Josep
  {Domingo-Ferrer}} {and} \bibinfo{person}{Maryline Laurent}} (Eds.).
  \bibinfo{publisher}{{Springer International Publishing}},
  \bibinfo{address}{{Cham}}, \bibinfo{pages}{172--187}.
\newblock
\showISBNx{978-3-031-13945-1}
\urldef\tempurl%
\url{https://doi.org/10.1007/978-3-031-13945-1_13}
\showDOI{\tempurl}


\bibitem[Broach et~al\mbox{.}(2012)]%
        {broach_where_2012}
\bibfield{author}{\bibinfo{person}{Joseph Broach}, \bibinfo{person}{Jennifer
  Dill}, {and} \bibinfo{person}{John Gliebe}.} \bibinfo{year}{2012}\natexlab{}.
\newblock \showarticletitle{Where Do Cyclists Ride? {{A}} Route Choice Model
  Developed with Revealed Preference {{GPS}} Data}.
\newblock \bibinfo{journal}{\emph{Transportation Research Part A: Policy and
  Practice}} \bibinfo{volume}{46}, \bibinfo{number}{10} (\bibinfo{date}{Dec.}
  \bibinfo{year}{2012}), \bibinfo{pages}{1730--1740}.
\newblock
\showISSN{0965-8564}
\urldef\tempurl%
\url{https://doi.org/10.1016/j.tra.2012.07.005}
\showDOI{\tempurl}


\bibitem[Choi et~al\mbox{.}(2021)]%
        {choi_trajgail_2021}
\bibfield{author}{\bibinfo{person}{Seongjin Choi}, \bibinfo{person}{Jiwon Kim},
  {and} \bibinfo{person}{Hwasoo Yeo}.} \bibinfo{year}{2021}\natexlab{}.
\newblock \showarticletitle{{{TrajGAIL}}: {{Generating}} Urban Vehicle
  Trajectories Using Generative Adversarial Imitation Learning}.
\newblock \bibinfo{journal}{\emph{Transportation Research Part C: Emerging
  Technologies}}  \bibinfo{volume}{128} (\bibinfo{year}{2021}),
  \bibinfo{pages}{103091}.
\newblock
\urldef\tempurl%
\url{https://doi.org/10.1016/j.trc.2021.103091}
\showDOI{\tempurl}


\bibitem[Dwork(2006)]%
        {dwork_differential_2006}
\bibfield{author}{\bibinfo{person}{Cynthia Dwork}.}
  \bibinfo{year}{2006}\natexlab{}.
\newblock \showarticletitle{Differential {{Privacy}}}. In
  \bibinfo{booktitle}{\emph{Automata, {{Languages}} and {{Programming}}}}
  \emph{(\bibinfo{series}{Lecture {{Notes}} in {{Computer Science}}})},
  \bibfield{editor}{\bibinfo{person}{Michele Bugliesi}, \bibinfo{person}{Bart
  Preneel}, \bibinfo{person}{Vladimiro Sassone}, {and} \bibinfo{person}{Ingo
  Wegener}} (Eds.). \bibinfo{publisher}{{Springer}}, \bibinfo{address}{{Berlin,
  Heidelberg}}, \bibinfo{pages}{1--12}.
\newblock
\showISBNx{978-3-540-35908-1}
\urldef\tempurl%
\url{https://doi.org/10.1007/11787006_1}
\showDOI{\tempurl}


\bibitem[Dwork and Roth(2014)]%
        {8187424}
\bibfield{author}{\bibinfo{person}{Cynthia Dwork} {and} \bibinfo{person}{Aaron
  Roth}.} \bibinfo{year}{2014}\natexlab{}.
\newblock \bibinfo{booktitle}{\emph{The Algorithmic Foundations of Differential
  Privacy}}.
\newblock


\bibitem[Gursoy et~al\mbox{.}(2018)]%
        {gursoy_utility-aware_2018}
\bibfield{author}{\bibinfo{person}{Mehmet~Emre Gursoy}, \bibinfo{person}{Ling
  Liu}, \bibinfo{person}{Stacey Truex}, \bibinfo{person}{Lei Yu}, {and}
  \bibinfo{person}{Wenqi Wei}.} \bibinfo{year}{2018}\natexlab{}.
\newblock \showarticletitle{Utility-{{Aware Synthesis}} of {{Differentially
  Private}} and {{Attack-Resilient Location Traces}}}. In
  \bibinfo{booktitle}{\emph{Proceedings of the 2018 {{ACM SIGSAC Conference}}
  on {{Computer}} and {{Communications Security}}}}. \bibinfo{pages}{196--211}.
\newblock
\urldef\tempurl%
\url{https://doi.org/10.1145/3243734.3243741}
\showDOI{\tempurl}


\bibitem[He et~al\mbox{.}(2015)]%
        {he_dpt_2015}
\bibfield{author}{\bibinfo{person}{Xi He}, \bibinfo{person}{Graham Cormode},
  \bibinfo{person}{Ashwin Machanavajjhala}, \bibinfo{person}{Cecilia~M.
  Procopiuc}, {and} \bibinfo{person}{Divesh Srivastava}.}
  \bibinfo{year}{2015}\natexlab{}.
\newblock \showarticletitle{{{DPT}}: {{Differentially Private Trajectory
  Synthesis Using Hierarchical Reference Systems}}}.
\newblock \bibinfo{journal}{\emph{Proceedings of the VLDB Endowment}}
  \bibinfo{volume}{8}, \bibinfo{number}{11} (\bibinfo{year}{2015}),
  \bibinfo{pages}{1154--1165}.
\newblock
\urldef\tempurl%
\url{https://doi.org/10.14778/2809974.2809978}
\showDOI{\tempurl}


\bibitem[Kapp(2022)]%
        {kapp_collection_2022}
\bibfield{author}{\bibinfo{person}{Alexandra Kapp}.}
  \bibinfo{year}{2022}\natexlab{}.
\newblock \showarticletitle{Collection, Usage and Privacy of Mobility Data in
  the Enterprise and Public Administrations}.
\newblock \bibinfo{journal}{\emph{Proceedings on Privacy Enhancing
  Technologies}} \bibinfo{volume}{2022}, \bibinfo{number}{4}
  (\bibinfo{date}{Oct.} \bibinfo{year}{2022}), \bibinfo{pages}{440--456}.
\newblock
\showISSN{2299-0984}
\urldef\tempurl%
\url{https://doi.org/10.56553/popets-2022-0117}
\showDOI{\tempurl}


\bibitem[Karakaya and Bermbach(2022)]%
        {karakaya_simra_2022}
\bibfield{author}{\bibinfo{person}{Ahmet-Serdar Karakaya} {and}
  \bibinfo{person}{David Bermbach}.} \bibinfo{year}{2022}\natexlab{}.
\newblock \bibinfo{title}{{SimRa Rides Berlin 10/21 - 09/22}}.
\newblock
\newblock
\urldef\tempurl%
\url{https://doi.org/10.14279/DEPOSITONCE-16439}
\showDOI{\tempurl}


\bibitem[Karakaya et~al\mbox{.}(2020)]%
        {karakaya_simra_2020-1}
\bibfield{author}{\bibinfo{person}{Ahmet-Serdar Karakaya},
  \bibinfo{person}{Jonathan Hasenburg}, {and} \bibinfo{person}{David
  Bermbach}.} \bibinfo{year}{2020}\natexlab{}.
\newblock \showarticletitle{{{SimRa}}: {{Using}} Crowdsourcing to Identify near
  Miss Hotspots in Bicycle Traffic}.
\newblock \bibinfo{journal}{\emph{Pervasive and Mobile Computing}}
  \bibinfo{volume}{67} (\bibinfo{date}{Sept.} \bibinfo{year}{2020}),
  \bibinfo{pages}{101197}.
\newblock
\showISSN{1574-1192}
\urldef\tempurl%
\url{https://doi.org/10.1016/j.pmcj.2020.101197}
\showDOI{\tempurl}


\bibitem[Lesty{\'a}n et~al\mbox{.}(2022)]%
        {lestyan_search_2022}
\bibfield{author}{\bibinfo{person}{Szilvia Lesty{\'a}n},
  \bibinfo{person}{Gergely {\'A}cs}, {and} \bibinfo{person}{Gergely
  Bicz{\'o}k}.} \bibinfo{year}{2022}\natexlab{}.
\newblock \showarticletitle{In Search of Lost Utility: {{Private}} Location
  Data}.
\newblock \bibinfo{journal}{\emph{Proceedings on Privacy Enhancing
  Technologies}} \bibinfo{volume}{2022}, \bibinfo{number}{3}
  (\bibinfo{year}{2022}), \bibinfo{pages}{354--372}.
\newblock
\urldef\tempurl%
\url{https://doi.org/10.56553/popets-2022-0076}
\showDOI{\tempurl}


\bibitem[Li et~al\mbox{.}(2010)]%
        {li_deriving_2010}
\bibfield{author}{\bibinfo{person}{Xiaojie Li}, \bibinfo{person}{Xiang Li},
  \bibinfo{person}{Daimin Tang}, {and} \bibinfo{person}{Xianrui Xu}.}
  \bibinfo{year}{2010}\natexlab{}.
\newblock \showarticletitle{Deriving Features of Traffic Flow around an
  Intersection from Trajectories of Vehicles}. In
  \bibinfo{booktitle}{\emph{2010 18th {{International Conference}} on
  {{Geoinformatics}}}}. \bibinfo{pages}{1--5}.
\newblock
\showISSN{2161-0258}
\urldef\tempurl%
\url{https://doi.org/10.1109/GEOINFORMATICS.2010.5567483}
\showDOI{\tempurl}


\bibitem[Lu et~al\mbox{.}(2018)]%
        {lu_understanding_2018}
\bibfield{author}{\bibinfo{person}{Wei Lu}, \bibinfo{person}{Darren~M. Scott},
  {and} \bibinfo{person}{Ron Dalumpines}.} \bibinfo{year}{2018}\natexlab{}.
\newblock \showarticletitle{Understanding Bike Share Cyclist Route Choice Using
  {{GPS}} Data: {{Comparing}} Dominant Routes and Shortest Paths}.
\newblock \bibinfo{journal}{\emph{Journal of Transport Geography}}
  \bibinfo{volume}{71} (\bibinfo{date}{July} \bibinfo{year}{2018}),
  \bibinfo{pages}{172--181}.
\newblock
\showISSN{0966-6923}
\urldef\tempurl%
\url{https://doi.org/10.1016/j.jtrangeo.2018.07.012}
\showDOI{\tempurl}


\bibitem[Luca et~al\mbox{.}(2021)]%
        {luca_survey_2021}
\bibfield{author}{\bibinfo{person}{Massimiliano Luca}, \bibinfo{person}{Gianni
  Barlacchi}, \bibinfo{person}{Bruno Lepri}, {and} \bibinfo{person}{Luca
  Pappalardo}.} \bibinfo{year}{2021}\natexlab{}.
\newblock \showarticletitle{A Survey on Deep Learning for Human Mobility}.
\newblock \bibinfo{journal}{\emph{ACM Computing Surveys (CSUR)}}
  \bibinfo{volume}{55}, \bibinfo{number}{1} (\bibinfo{year}{2021}),
  \bibinfo{pages}{1--44}.
\newblock


\bibitem[Luxen and Vetter(2011)]%
        {luxen-vetter-2011}
\bibfield{author}{\bibinfo{person}{Dennis Luxen} {and}
  \bibinfo{person}{Christian Vetter}.} \bibinfo{year}{2011}\natexlab{}.
\newblock \showarticletitle{Real-Time Routing with {{OpenStreetMap}} Data}. In
  \bibinfo{booktitle}{\emph{Proceedings of the 19th {{ACM SIGSPATIAL}}
  International Conference on Advances in Geographic Information Systems}}
  \emph{(\bibinfo{series}{{{GIS}} '11})}. \bibinfo{publisher}{{ACM}},
  \bibinfo{address}{{New York, NY, USA}}, \bibinfo{pages}{513--516}.
\newblock
\showISBNx{978-1-4503-1031-4}
\urldef\tempurl%
\url{https://doi.org/10.1145/2093973.2094062}
\showDOI{\tempurl}


\bibitem[Prato(2009)]%
        {prato_route_2009}
\bibfield{author}{\bibinfo{person}{Carlo~Giacomo Prato}.}
  \bibinfo{year}{2009}\natexlab{}.
\newblock \showarticletitle{Route Choice Modeling: Past, Present and Future
  Research Directions}.
\newblock \bibinfo{journal}{\emph{Journal of Choice Modelling}}
  \bibinfo{volume}{2}, \bibinfo{number}{1} (\bibinfo{date}{Jan.}
  \bibinfo{year}{2009}), \bibinfo{pages}{65--100}.
\newblock
\showISSN{1755-5345}


\bibitem[Stadler and Troncoso(2022)]%
        {Stadler:298904}
\bibfield{author}{\bibinfo{person}{Theresa Stadler} {and}
  \bibinfo{person}{Carmela Troncoso}.} \bibinfo{year}{2022}\natexlab{}.
\newblock \showarticletitle{Why the Search for a Privacy-Preserving Data
  Sharing Mechanism Is Failing}.
\newblock \bibinfo{journal}{\emph{Nature Computational Science}}
  \bibinfo{volume}{2}, \bibinfo{number}{4} (\bibinfo{year}{2022}),
  \bibinfo{pages}{208--210}.
\newblock
\urldef\tempurl%
\url{https://doi.org/10.1038/s43588-022-00236-x}
\showDOI{\tempurl}


\bibitem[Sun et~al\mbox{.}(2023)]%
        {sun_synthesizing_2023}
\bibfield{author}{\bibinfo{person}{Xinyue Sun}, \bibinfo{person}{Qingqing Ye},
  \bibinfo{person}{Haibo Hu}, \bibinfo{person}{Yuandong Wang},
  \bibinfo{person}{Kai Huang}, \bibinfo{person}{Tianyu Wo}, {and}
  \bibinfo{person}{Jie Xu}.} \bibinfo{year}{2023}\natexlab{}.
\newblock \showarticletitle{Synthesizing {{Realistic Trajectory Data With
  Differential Privacy}}}.
\newblock \bibinfo{journal}{\emph{IEEE Transactions on Intelligent
  Transportation Systems}} (\bibinfo{year}{2023}), \bibinfo{pages}{1--14}.
\newblock
\showISSN{1558-0016}


\bibitem[Wang et~al\mbox{.}(2023)]%
        {wang2023privtrace}
\bibfield{author}{\bibinfo{person}{Haiming Wang}, \bibinfo{person}{Zhikun
  Zhang}, \bibinfo{person}{Tianhao Wang}, \bibinfo{person}{Shibo He},
  \bibinfo{person}{Michael Backes}, \bibinfo{person}{Jiming Chen}, {and}
  \bibinfo{person}{Yang Zhang}.} \bibinfo{year}{2023}\natexlab{}.
\newblock \showarticletitle{{{PrivTrace}}: {{Differentially}} Private
  Trajectory Synthesis by Adaptive Markov Model}. In
  \bibinfo{booktitle}{\emph{{{USENIX}} Security Symposium 2023}}.
\newblock


\bibitem[Wei et~al\mbox{.}(2020)]%
        {wei_survey_2020}
\bibfield{author}{\bibinfo{person}{Hua Wei}, \bibinfo{person}{Guanjie Zheng},
  \bibinfo{person}{Vikash Gayah}, {and} \bibinfo{person}{Zhenhui Li}.}
  \bibinfo{year}{2020}\natexlab{}.
\newblock \showarticletitle{A {{Survey}} on {{Traffic Signal Control
  Methods}}}.
\newblock  (\bibinfo{date}{Jan.} \bibinfo{year}{2020}).
\newblock
\showeprint[arxiv]{1904.08117}~[cs, stat]


\end{thebibliography}
